\documentclass[sigconf]{acmart}

\usepackage{makecell}
\usepackage{booktabs}
\usepackage{pifont}
\usepackage{mathpartir}
\usepackage{longtable}
\usepackage{multicol}
\usepackage{amsmath}
\usepackage{amsthm}
\usepackage{algorithm}
\usepackage[noend]{algpseudocode}
\usepackage{xspace}
\usepackage{lipsum}
\usepackage{float}
\usepackage[english]{babel}
\usepackage{hyperref}
\usepackage{xcolor,colortbl}
\definecolor{RoyalBlue}{RGB}{65, 105, 225}
\definecolor{Orange}{RGB}{255, 165, 0}
\usepackage{listings}

\usepackage{listings, xcolor}

\definecolor{verylightgray}{rgb}{.97,.97,.97}

\lstdefinelanguage{Solidity}{
	keywords=[1]{anonymous, assembly, assert, balance, break, call, callcode, case, catch, class, constant, continue, constructor, contract, debugger, default, delegatecall, delete, do, else, emit, event, experimental, export, external, false, finally, for, function, gas, if, implements, import, in, indexed, instanceof, interface, internal, is, length, library, log0, log1, log2, log3, log4, memory, modifier, new, payable, pragma, private, protected, public, pure, push, require, return, returns, revert, selfdestruct, send, solidity, storage, struct, suicide, super, switch, then, this, throw, true, try, typeof, using, value, view, while, with, addmod, ecrecover, keccak256, mulmod, ripemd160, sha256, sha3}, 
	keywordstyle=[1]\color{blue}\bfseries,
	keywords=[2]{address, bool, byte, bytes, bytes1, bytes2, bytes3, bytes4, bytes5, bytes6, bytes7, bytes8, bytes9, bytes10, bytes11, bytes12, bytes13, bytes14, bytes15, bytes16, bytes17, bytes18, bytes19, bytes20, bytes21, bytes22, bytes23, bytes24, bytes25, bytes26, bytes27, bytes28, bytes29, bytes30, bytes31, bytes32, enum, int, int8, int16, int24, int32, int40, int48, int56, int64, int72, int80, int88, int96, int104, int112, int120, int128, int136, int144, int152, int160, int168, int176, int184, int192, int200, int208, int216, int224, int232, int240, int248, int256, mapping, string, uint, uint8, uint16, uint24, uint32, uint40, uint48, uint56, uint64, uint72, uint80, uint88, uint96, uint104, uint112, uint120, uint128, uint136, uint144, uint152, uint160, uint168, uint176, uint184, uint192, uint200, uint208, uint216, uint224, uint232, uint240, uint248, uint256, var, void, ether, gwei, finney, szabo, wei, days, hours, minutes, seconds, weeks, years},	
	keywordstyle=[2]\color{teal}\bfseries,
	keywords=[3]{block, blockhash, coinbase, difficulty, gaslimit, number, timestamp, msg, data, gas, sender, sig, value, now, tx, gasprice, origin},	
	keywordstyle=[3]\color{violet}\bfseries,
	identifierstyle=\color{black},
	sensitive=true,
	comment=[l]{//},
	morecomment=[s]{/*}{*/},
	commentstyle=\color{gray}\ttfamily,
	stringstyle=\color{red}\ttfamily,
	morestring=[b]',
	morestring=[b]"
}

\usepackage{multirow}
\usepackage{syntax}
\usepackage{enumitem}
\usepackage{comment}
\newcommand{\ballnumber}[1]{\tikz[baseline=(myanchor.base)] \node[circle,fill=.,inner sep=1pt] (myanchor) {\color{-.}\bfseries\footnotesize #1};}
\usepackage{tikz}
\usetikzlibrary{positioning,shapes,arrows}
\usetikzlibrary{patterns}
\usepackage{pgfplots}
\usepackage{pgf-pie}
\usepackage{rotating}
\usepackage{textcomp}

\usepackage{bbding}
\usepackage{subcaption}
\usepackage{soul}

\usepackage[nameinlink]{cleveref}
\crefname{section}{Section}{Sections}
\crefname{figure}{Figure}{Figures}
\crefname{table}{Table}{Tables}
\crefname{algorithm}{Algorithm}{Algorithms}
\crefname{section}{Section}{Sections}
\crefname{proposition}{Proposition}{Propositions}
\crefname{assumption}{Assumption}{Assumptions}
\crefname{equation}{Equation}{Equations}
\crefname{remark}{Remark}{Remarks}
\crefname{example}{Example}{Examples}
\crefname{appendix}{Appendix}{Appendices}
\crefname{chapter}{Chapter}{Chapters}

\AtBeginDocument{%
  }

\copyrightyear{2024}
\acmYear{2024}
\setcopyright{rightsretained}
\acmConference[CCS '24]{Proceedings of the 2024 ACM SIGSAC Conference on Computer and Communications Security}{October 14--18, 2024}{Salt Lake City, UT, USA}
\acmBooktitle{Proceedings of the 2024 ACM SIGSAC Conference on Computer and Communications Security (CCS '24), October 14--18, 2024, Salt Lake City, UT, USA}
\acmDOI{10.1145/3658644.3690293}
\acmISBN{979-8-4007-0636-3/24/10}

\begin{document}

\title{FORAY: Towards Effective Attack Synthesis against Deep Logical Vulnerabilities in DeFi Protocols}


\author{Hongbo Wen}
\email{hongbowen@ucsb.edu}
\orcid{0000-0003-3517-445X}
\affiliation{%
  \institution{University of California, Santa Barbara}
  \city{Santa Barbara}
  \state{California}
  \country{USA}
}

\author{Hanzhi Liu}
\email{hanzhi@ucsb.edu}
\orcid{0000-0002-7027-8302}
\affiliation{%
  \institution{University of California, Santa Barbara}
  \city{Santa Barbara}
  \state{California}
  \country{USA}
}

\author{Jiaxin Song}
\email{jiaxins8@illinois.edu}
\orcid{0000-0002-4847-9183}
\affiliation{%
  \institution{University of Illinois Urbana-Champaign}
  \city{Champaign}
  \state{Illinois}
  \country{USA}
}

\author{Yanju Chen}
\email{yanju@cs.ucsb.edu}
\orcid{0000-0002-6494-3126}
\affiliation{%
  \institution{University of California, Santa Barbara}
  \city{Santa Barbara}
  \state{California}
  \country{USA}
}


\author{Wenbo Guo}
\email{henrygwb@ucsb.edu}
\orcid{0000-0002-6890-4503}
\affiliation{%
  \institution{University of California, Santa Barbara}
  \city{Santa Barbara}
  \state{California}
  \country{USA}
}

\author{Yu Feng}
\email{yufeng@cs.ucsb.edu}
\orcid{0000-0003-1000-1229}
\affiliation{%
  \institution{University of California, Santa Barbara}
  \city{Santa Barbara}
  \state{California}
  \country{USA}
}

\renewcommand{\shortauthors}{Hongbo Wen et al.}

\begin{abstract}
  A clear and well-documented \LaTeX\ document is presented as an
  article formatted for publication by ACM in a conference proceedings
  or journal publication. Based on the ``acmart'' document class, this
  article presents and explains many of the common variations, as well
  as many of the formatting elements an author may use in the
  preparation of the documentation of their work.
\end{abstract}



\newcommand{\yanju}[1]{{{\color{orange}{{(Yanju: #1)\xspace}}}}}
\newcommand{\rev}[1]{{{\color{orange}{{#1\xspace}}}}}
\newcommand{\hongbo}[1]{{{\color{purple}{{(Hongbo: #1)\xspace}}}}}
\newcommand{\yu}[1]{{{\color{red}{{(Yu: #1)\xspace}}}}}
\newcommand{\wenbo}[1]{{{\color{blue}{{(Wenbo: #1)\xspace}}}}}
\newcommand{\revpoint}[1]{{{\color{red}{{(Shepherd: #1)\xspace}}}}}
\newcommand{\smartparagraph}[1]{\noindent{\bf #1}\ }

\newcommand{\tool}{{\textsc{Foray}}\xspace}
\newcommand{\toolab}{{\textsc{Foray}$^o$}\xspace}
\newcommand{\hole}{{\diamond}\xspace}
\newcommand{\void}{\epsilon\xspace}
\newcommand{\constraint}{\Phi\xspace}
\newcommand{\attacker}{\varmathbb{A}\xspace}
\newcommand{\fundom}{\varmathbb{F}\xspace}
\newcommand{\tfg}{\varmathbb{G}\xspace}
\newcommand{\afl}{\varmathbb{P}\xspace}
\newcommand{\token}{\varmathbb{T}\xspace}
\newcommand{\edgedom}{\varmathbb{E}\xspace}
\newcommand{\markdom}{\varmathbb{M}\xspace}
\newcommand{\reach}{\mathcal{R}}
\newcommand{\tuple}[1]{{\langle #1 \rangle}}
\newcommand{\cststore}{\Omega}
\newcommand{\muc}{\kappa}
\newcommand{\halmos}{{\textsc{Halmos}}\xspace}
\newcommand{\ityfuzz}{{\textsc{ItyFuzz}}\xspace}

\newcommand{\inlinecode}[1]{{\texttt{\small #1\xspace}}}

\newcommand{\reducedstrut}{\vrule width 0pt height .1\ht\strutbox depth .1\dp\strutbox\relax}
\newcommand{\code}[1]{%
  \begingroup
  \setlength{\fboxsep}{0pt}%
  \colorbox{RoyalBlue!15}{\reducedstrut\texttt{\small #1}\/}%
  \endgroup
}
\newcommand{\irule}[1]{%
  \begingroup
  \setlength{\fboxsep}{0pt}%
  \colorbox{Orange!20}{\reducedstrut\textbf{\sf\small #1}\/}%
  \endgroup
}

\algnewcommand\algorithmicforeach{\textbf{for each}}
\algdef{S}[FOR]{ForEach}[1]{\algorithmicforeach\ #1\ \algorithmicdo}

\begin{abstract}


Blockchain adoption has surged with the rise of Decentralized Finance (DeFi) applications. However, the significant value of digital assets managed by DeFi protocols makes them prime targets for attacks. Current smart contract vulnerability detection tools struggle with DeFi protocols due to deep logical bugs arising from complex financial interactions between multiple smart contracts. These tools primarily analyze individual contracts and resort to brute-force methods for DeFi protocols crossing numerous smart contracts, leading to inefficiency.

We introduce \tool, a highly effective attack synthesis framework against deep logical bugs in DeFi protocols. 
\tool proposes a novel attack sketch generation and completion framework.
Specifically, instead of treating DeFis as regular programs, we design a domain-specific language (DSL) to lift the low-level smart contracts into their high-level financial operations.
Based on our DSL, we first compile a given DeFi protocol into a token flow graph, our graphical representation of DeFi protocols. 
Then, we design an efficient sketch generation method to synthesize attack sketches for a certain attack goal (e.g., price manipulation, arbitrage, etc.).
This algorithm strategically identifies candidate sketches by finding reachable paths in \emph{Token Flow Graph} (TFG), which is much more efficient than random enumeration. 
For each candidate sketch written in our DSL, \tool designs a domain-specific symbolic compilation to compile it into SMT constraints.
Our compilation simplifies the constraints by removing redundant smart contract semantics. 
It maintains the usability of symbolic compilation, yet scales to problems orders of magnitude larger. 
Finally, the candidates are completed via existing solvers and are transformed into concrete attacks via direct syntax transformation.
Through extensive experiments on real-world security incidents, we demonstrate that \tool significantly outperforms \halmos and \ityfuzz, the state-of-the-art (SOTA) tools for smart contract vulnerability detection, in both effectiveness and efficiency.
Specifically, out of 34 benchmark DeFi logical bugs that happened in the last two years, \tool synthesizes 27 attacks, whereas \ityfuzz and \halmos only synthesize 11 and 3, respectively.
Furthermore, \tool also finds \textit{ten} zero-day vulnerabilities in the BNB chain.
Finally, we demonstrate the effectiveness of our key components and \tool's capability of avoiding false positives.

\end{abstract}

\maketitle

\section{Introduction}
\label{sec:intro}

Decentralized Finance (DeFi) applications have driven a surge in blockchain adoption by offering real-world financial services like lending, borrowing, and trading on blockchain networks. 
This has brought in a broader user base and increased interest in blockchain technology, with a total funding amount of more than \$90 billion locked in DeFi applications as of March 2023~\cite{defillama}. 
Nonetheless, the substantial value of digital assets under the management of DeFis renders them an enticing target for potential attacks. 
For instance, the recent price manipulation vulnerability~\cite{priceattack1,priceattack2,priceattack3} allows malicious actors to induce DeFi protocols (a set of smart contracts that realize a certain financial model) to execute transactions that are detrimental to user's funds. 
Furthermore, attackers can manipulate DeFi protocols to instigate exchanges from lower-valued assets to higher-valued ones or to secure significant loans, often using low-value assets as collateral. 
This manipulation is achieved by tampering with the circulation of tokens, thus influencing token prices in the process. 
Statistics from the incomplete hack event database~\cite{DeFiHackLabs} show that attacks exploiting logical flaws of \emph{financial models} behind DeFis (denoted as deep logical bugs) have resulted in a cumulative loss of up to \$200 million over the past two years.

Improving the robustness of DeFi protocols is thus a pressing concern and there has been a flurry of research~\cite{oyente,securify,mythril,summarybased,sailfish2022} in the past few years. 
However, the majority of current detection tools primarily concentrate on code vulnerabilities of a single contract, such as re-entrancy, integer overflow, access control, etc. 
Therefore, it is unsurprised that these tools cannot be employed effectively to identify DeFi attacks stemming from logic flaws. 
The complexity of multiple contracts in DeFi and their interactions dramatically increase the search space that goes beyond the capability of existing analyzers. 
To make things even worse, the smart contracts in DeFis are immutable -- once they are deployed, fixing their bugs is extremely difficult due to the design of the consensus protocol.

We introduce \tool, a synthesizer for automatically generating exploits against deep logical bugs in DeFi protocols. 
\tool introduces an attack sketch generation and completion framework. 
It first generates incomplete attack sketches written in our DSL. 
Then, it leverages our proposed \emph{domain-specific symbolic compilation} approach to compile the attack sketches with logical holes into constraints that can be solved by off-the-shelf solvers.
Finally, it fills the holes with a SOTA solver and transforms the complete sketches into concrete attacks through a direct syntax transformation. 

The key technical challenges are two-fold. 
First, existing tools cannot strategically generate sketches for DeFi beyond random enumeration. 
Second, current symbolic compilation tools treat DeFi as a collection of regular smart contracts, disregarding the high-level~\emph{financial models} in DeFi protocols. 
To mitigate the first challenge, given a DeFi protocol, \tool first compiles it into a \emph{Token Flow Graph} (TFG), our proposed high-level semantic representation for DeFi protocols. 
Here, nodes represent different tokens (USDC, WETH, USDT, etc.) and edges are labeled with constructs from \tool's \emph{abstract financial language}, which provides high-level operators (e.g., lend/borrow/pay/swap) over financial assets. 
Now, given a particular attack goal (e.g., price manipulation, arbitrage, etc.) in the form of a logical formula, \tool models the attack sketch generation as a reachability problem in TFG.
Instead of random enumeration, \tool devises an effective \emph{sketch generation} algorithm that strategically enumerates relevant attack sketches using a \emph{type-directed} graph reachability algorithm over the TFG.

To tackle the second challenge, \tool employs a domain-specific symbolic compilation strategy, which maintains the usability of symbolic compilation, yet scales to problems orders of magnitude larger. 
For each candidate attack sketch, \tool leverages the \emph{abstract semantics} of our proposed DSL to compile possible completions of the sketch into SMT constraints that can be efficiently solved by off-the-shelf solvers~\cite{z3}. 
Here, our domain-specific symbolic compilation can filter out low-level smart contract semantics and thus significantly simplify the constraints. 
Because both our \emph{sketch generation} and \emph{sketch completion} overapproximate the concrete semantics of DeFis, \tool may generate spurious attacks that fail to achieve the goal. 
We mitigate this problem by incorporating a CEGIS (Counter Example-Guided Inductive Synthesis) loop that iteratively adds the root cause of the failed attempt to \tool's \emph{knowledge base}, which avoids similar mistakes in future iterations. 

We implement \tool and compare it against \halmos~\cite{halmos} and \ityfuzz~\cite{ityfuzz}, the state-of-the-art tools for analyzing smart contracts and DeFi protocols.  
Our experiment shows that our tool is efficient and effective. 
On the set of 34 security incidents in the past two years, \tool manages to synthesize attacks for 79\% of the benchmarks with an average synthesis time of 105.9 seconds.
On the other hand, Halmos can only solve 10\% of the benchmarks with an average running time of 8085.0 seconds, which demonstrates that \tool's domain-specific symbolic compilation accelerates synthesis several orders of magnitude compared to the general-purpose compilation to an SMT solver. 
Furthermore, we also apply \tool to DeFi protocols on the BNB chain~\cite{bnbchain2023whitepaper} and uncover \emph{ten} zero-day vulnerabilities with concrete attacks.
Finally, we verify the effectiveness of sketch generation and completion through an ablation study and demonstrate \tool's capability in alleviating false positives. 
Overall, \tool provides a novel attack synthesis technique against various types of deep logical bugs in DeFis protocols.

In summary, this paper makes the following contributions:
\begin{itemize}
    
    \item We propose \emph{Abstract Financial Language}, a DSL that describes high-level financial operators in DeFis. We also design \emph{Token Flow Graph}, a semantic representation that summarizes the financial model of a DeFi protocol. 
    
    \item We propose an effective CEGIS framework for DeFi attack synthesis. In particular, our sketch generation leverages a type-directed graph reachability over a token flow graph and our sketch completion designs a domain-specific symbolic compilation strategy that results in easy-to-solve constraints.

    \item We implement the proposed ideas in a tool called \tool
    and demonstrate that it achieves several orders of magnitude speed-up compared to general-purpose symbolic compilation. Furthermore, \tool not only generated 80\% security incidents in the past two years (2022-2023) but also detected ten zero-day DeFi vulnerabilities from popular blockchains.
\end{itemize}
\section{Background}
\label{sec:background}

\subsection{Blockchain basis.}
\label{subsec:2.1}

\smartparagraph{Ethereum.}
Blockchain functions as a decentralized record-keeping platform that chronicles and disseminates transaction data among multiple users. 
It is an expand-only chain of interconnected blocks, managed by a consensus mechanism, where each block contains a collection of transactions. 
Among various blockchain systems, Ethereum~\cite{wood2014ethereum} is the first blockchain capable of storing, managing, and running Turing-complete scripts, termed~\textit{smart contracts}. 
Ethereum operates on a comprehensive state system updated via transaction execution. 
The transactions are initiated by and received by users through their accounts.  
Ethereum has two principal types of accounts: those owned by users and those governed by smart contracts, each associated with a distinct \textit{address}. 
Besides making transactions, users can also develop customized smart contracts that are programmed to execute transactions autonomously. 

\smartparagraph{Tokens and cryptocurrencies.}
Among different types of smart contracts, Tokens are a specific type that represents cryptocurrencies. 
Each Token contract must adhere to standardized interfaces like ERC20~\cite{ERC20TokenStandard2015}, ERC721~\cite{ERC721TokenStandard2018}, and ERC1155~\cite{ERC1155MultiTokenStandard2018}, which define how users interact with the corresponding token. 
For Ethereum, ERC20 is the most widely adopted interface.
To tether the value of cryptocurrencies to fiat currency, \textit{stablecoins}—like USDT~\cite{Tether2023}, which is implemented as an ERC20 token--have been created. 
They are pegged to the dollar reserves held by the issuer, providing a stable reference point for the value of other cryptocurrencies.

\subsection{Decentralized Finance (DeFi)}
\label{subsec:2.2}

Decentralized Finance (DeFi) refers to a set of financial applications built on blockchain technology.
They aim to recreate traditional financial systems, such as banking and lending, but without the need for intermediaries like banks or brokers. 
Instead, each DeFi service is implemented as a protocol that amalgamates various smart contracts.
Users access a DeFi service by engaging with the corresponding protocol through transactions.
According to a recent survey~\cite{DefiPrime2023Ethereum}, over 200 DeFi applications have been launched on the Ethereum platform. 
Here we list three major DeFi applications:

\smartparagraph{Lending.} platforms (such as Aave~\cite{aave}, MakerDAO~\cite{makerdao}) enable users to obtain on-chain cryptocurrencies as loans by depositing collateral into the system. 
The interest rates for borrowing are set by the DeFi protocols while maintaining transparency for users. 
As market conditions fluctuate, the collateral's value may fall below or rise above a certain threshold. 
When this happens, either the application or other users can liquidate or sell the collateral to gain profits.

\smartparagraph{Flash loans.} (e.g., dYdX~\cite{dydx}, Uniswap~\cite{uniswap}) represent a collateral-free borrowing model. 
This enables the borrower to run custom code through a callback function, with the stipulation that the loan must be repaid within the same transaction. 
If the borrower fails to return the loaned tokens, the lender will automatically reverse the lending transaction, ensuring that no permanent changes (to storage variables) are made by this transaction.

\smartparagraph{Decentralized exchanges (DEXs).} function as cryptocurrency exchanges that enable users to trade various tokens through direct interaction with smart contracts. 
These platforms incentivize users to deposit pairs or multiple tokens into a liquidity pool. 
As long as the pool maintains sufficient token volume, users can execute token swaps within it. 
The exchange rate for these trades is determined autonomously by the application's built-in pricing algorithm.
Popular DEXs protocols include 1inch~\cite{1inch}, PancakeSwap~\cite{pancake}.

\smartparagraph{DeFi vulnerabilities.}
At a high level, there are two types of vulnerabilities in DeFi protocols.
The first type refers to vulnerabilities in individual smart contracts, including assertion failures, arbitrary writes, control-flow hijacking, etc (denoted as common vulnerabilities).
These vulnerabilities are similar to traditional software security bugs and are possible to be automatically detected by analyzing the smart contract code.
As discussed in \cref{sec:related}, existing research works propose a number of tools that utilize static and dynamic program analysis to automatically identify such vulnerabilities.
The second type of vulnerability exploits logical flaws in a DeFi protocol, which we refer to as~\textbf{deep logical bugs} in this paper. 
As demonstrated in \cref{sec:motivating}, these deep logical bugs exploit public functions across multiple smart contracts within the DeFi protocol to maliciously increase an attacker's profits. 
Identifying such vulnerabilities is extremely challenging because it requires a deep understanding of the semantics and business logic of the DeFi protocol, as well as the composition of transaction sequences. 
As shown in recent studies~\cite{zhou2023sok,zhang2023demystifying}, most existing tools designed for smart contract vulnerabilities fail to detect deep logical bugs. 
\newcommand{\examplename}{MUMUG\xspace}
\newcommand{\exampletoken}{MU\xspace}
\newcommand{\examplestable}{USDCe\xspace}
\begin{figure}[t]
    \centering
    \includegraphics[width=\linewidth]{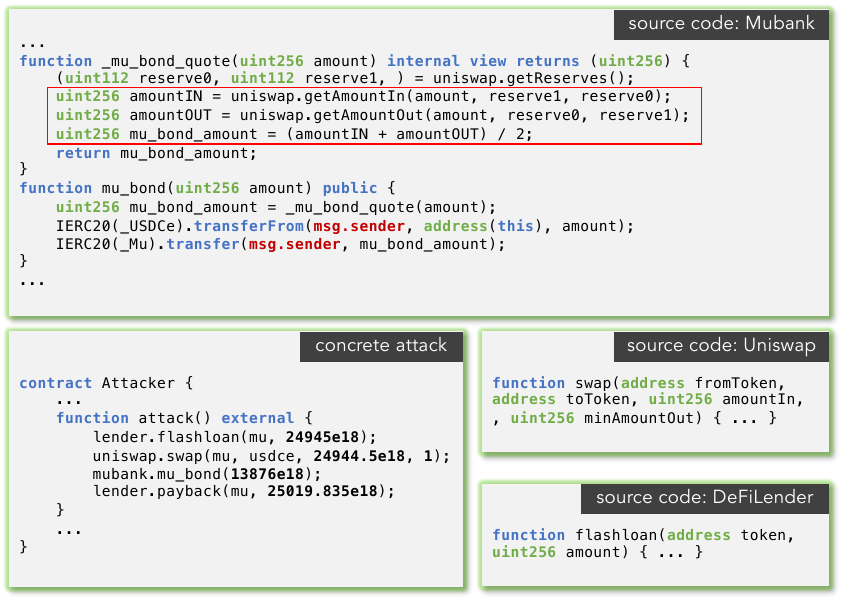}
    \vspace{-6mm}
    \caption{Illustration of \examplename and a concrete exploit against it.
    \code{IERC20().transferFrom} and \code{IERC20().transfer} are standard APIs that enable the withdraw and deposit of tokens for one address.
    \code{uniswap.getAmountIn} and \code{uniswap.getAmountOut} are \code{uniswap} APIs that calculate the required amount to swap one type of token for another based on their current reserves. 
    }
    \label{fig:motivating}
    \vspace{-5mm}
\end{figure}

\section{Problem Definition and Existing Solutions}
\label{sec:motivating}

In this section, we begin by specifying our problem scopes and demonstrating a deep logical bug of a simplified DeFi protocol, \examplename, which was hacked in 2022, resulting in the loss of nearly all its stablecoins. 
Then, we formally define DeFi attack synthesis and discuss the limitations of existing solutions.

\subsection{Problem Scope and Technical Challenges}
\label{subsec:3.1}

\smartparagraph{Threat model.} 
Our goal is to detect \emph{deep logical bugs} in a DeFi protocol by synthesizing a sequence of attack transactions that can exploit the DeFi protocol to gain profits maliciously. 
We assume an entirely trustless setup where an attacker can access all public information, including but not limited to on-chain blockchain states and the victim contracts' source code.
For contracts with only bytecodes, their source code can be obtained via reserve engineering, which is not our focus.  
Additionally, beyond directly interacting with the victim contracts, we assume the attacker can deploy their own contract, which can invoke public transactions of the target victim contracts (either directly or through callbacks). 
The attacker's goal is to synthesize a sequence of transactions that exploit the logical flaws in the target DeFi protocol to gain extra profit.  
We do not consider the common vulnerabilities.

\smartparagraph{\examplename protocol and an attack.}  
As shown in \cref{fig:motivating}, the protocol is composed of three key smart contracts: a) \code{DeFiLender} provides the \code{flashloan} function to enable the borrower to get tokens without collateral; 
b) \code{Mubank} with two functionalities.
The internal function (\code{_mu_bond_quote}) manages the sale and price of \exampletoken tokens based on the current reserves of \exampletoken and \examplestable.
It takes as input the amount of \examplestable and outputs the corresponding amount of \exampletoken in the same value.
The public function (\code{mu_bond}) enables users to withdraw \exampletoken by providing the same value of \examplestable determined by \code{_mu_bond_quote}. 
c) \code{Uniswap} is a popular protocol, which defines swap pairs for two types of tokens (e.g., \exampletoken and \examplestable).
Its \code{swap} function enables users to exchange tokens in a swap pair.
These three smart contracts define the \examplename DeFi protocol where benign users can borrow, withdraw, and exchange \exampletoken with \examplestable.

The susceptibility of \examplename lies in the pricing mechanism in the \code{Mubank} contract (highlighted in \cref{fig:motivating}). 
Given that the price of \exampletoken is determined by the reserve of \examplestable and \exampletoken within the swap pair. 
A significant fluctuation in the reserve level can result in an unexpectedly high volume of \exampletoken tokens and significantly lower its price.
An attack can leverage the price difference to withdraw the MU bank's stablecoins.
A concrete attack is shown in \cref{fig:motivating}. 
\ballnumber{1} Borrow a huge amount of \exampletoken tokens through the flashloan function in \code{DeFiLender}.
\ballnumber{2} Swap those \exampletoken tokens to a large amount of \examplestable at the swap pair. 
This will dramatically increase the reserve balance ratio of \exampletoken to \examplestable, devaluing the \exampletoken.
\ballnumber{3} Leverage the abnormal reserve balance ratio to swap a tiny amount of \examplestable for a huge amount of \exampletoken tokens at \code{MuBank}.
\ballnumber{4} Pay \exampletoken tokens back to the flash loan lender, keeping the majority of \examplestable acquired at step \ballnumber{2} as the profit.
Through this process, the attacker harvested approximately 57,660 \examplestable from the \code{MuBank}.

\smartparagraph{Formal definition of attack synthesis for deep logical bugs.}
Automatic attack synthesis in DeFi is equivalent to finding a sequence of function calls that exploit deep logical bugs of the DeFi protocol.
This can be formally defined as 
%
\begin{definition}[DeFi Attack Synthesis]
\label{def:3.1}
An attack synthesis for a DeFi protocol $D$ is a tuple $(L, S_0, \psi)$, where $L$ is the domain-specific language (DSL) for constructing the attack program.
For instance, a list of public functions is provided by the victim DeFi protocol.
$S_0$ is the initial and public blockchain state, and $\psi$ is the attack goal written in a logical formula.
DeFi attack synthesis is equivalent to finding an attack program $P$ written in DSL $L$, such that $P(S_0) \models \psi$ where $P(S_0)$ denotes the resulting state after executing $P$ on $S_0$. 
\end{definition}

\smartparagraph{Technical challenges.}
It is extremely challenging for the following two reasons. 
First, the search space is huge. 
In fact, \examplename protocol itself contains 26 public functions and the attackers can freely call public functions of other smart contracts (e.g., \code{uniswap.swap}). 
Even when we constrain the length of the function call sequence, the number of possible sequences is still extremely huge.
Searching a malicious function call sequence in such a huge search space is equivalent to finding a needle in a haystack.
Second, smart contracts and DeFi protocols have complicated semantics.
This imposes extra challenges to automatically represent a DeFi protocol with logical representations, making it hard to reason and synthesize attacks. 

\subsection{Existing Solutions and Limitations}
\label{subsec:3.2}

While attack synthesis is a novel concept in DeFi, it has been explored in traditional software security and program synthesis domains~\cite{morpheus,neo,summarybased}. 
Without any heavy customization, we can draw inspiration from traditional program synthesis and try to solve the problem with the following two solutions.

\smartparagraph{Static analysis and symbolic execution based-sketch generation and completion.}
Given that synthesizing the entire attack program from scratch is unlikely to scale, existing works in program synthesis usually decompose the synthesis into two phases~\textit{sketch generation} and~\textit{sketch completion}.
Here, an attack sketch refers to a sequence of actions, where each action is a function call to a certain smart contract.
Formally, we define an attack sketch $\tilde{P}$ as a sequence of invocations to constructs in $L$ where some of the constructs contain holes or symbolic variables yet to fill in.

To avoid exploring sketches doomed to fail, existing approaches typically leverage the \textit{abstract semantics} to only preserve sketches whose abstract semantics are \textit{consistent} with the attack goal $\psi$, $\tilde{P}(S_0) \Rightarrow \psi$, where $\tilde{P}(S_0)$ corresponds to the program state by \textit{abstractly} evaluating the sketch $\tilde{P}$ on $S_0$.
Then, the sketch completion step fills in the holes $\hole$ in each feasible sketch $\tilde{P}$ ($P = \tilde{P}[\mu/\hole]$) with language constructs $\mu$ in $L$ using symbolic execution, such that $P(S_0) \models \psi$. Each hole in \tool represents a function parameter. By resolving these parameters, the attack sketch is transformed into a concrete program and its execution result satisfies the attack goal.

The main challenges of this solution are as follows: 
First, there are no existing tools in DeFi that can effectively generate feasible attack sketches. 
The only way is to randomly select and combine function calls, which is extremely inefficient given the huge search space. 
Second, due to the complex semantics of DeFi protocols, the corresponding symbolic constraints of attack goals are intricate and often beyond the reasoning capacity of SOTA SMT solvers.
Specifically, to verify $P(S_0) \models \psi$, existing approaches have to reason about program $P$ by faithfully following the operational semantics of the host language $L$, which contains language features (e.g., gas consumption and memory models in Solidity.) and low-level details irrelevant to the synthesis goal.
As demonstrated in \cref{sec:eval}, it is extremely difficult for Halmos~\cite{halmos}, a SOTA symbolic testing tool for Ethereum smart contracts~\cite{summarybased,manticore,mythril}, to solve the constraints for common attacks within a feasible time limit. 

\smartparagraph{Fuzzing.}
SOTA fuzzers (e.g., ItyFuzz~\cite{ityfuzz} and Smartian~\cite{SMARTIAN}) in smart contracts support synthesizing sequences of actions that lead to vulnerabilities (violation of DeFi protocol). 
Fuzzing is more computationally efficient than symbolic execution-based solutions but it relies more on random generations and mutations.
In addition, due to DeFis' complex semantics, existing fuzzers do not have fitness functions or testing oracles that correspond to specific attack goals and thus cannot provide proper feedback signals of whether the current input is valid, making it even more difficult to find valid attacks through random mutations.   

Note that as discussed in \cref{sec:related}, there are some recent tools for automatically detecting DeFi protocol vulnerabilities. 
Most tools rely on summarizing attack patterns from past attack incidents and thus are hindered by the limited scope of these patterns. 
They can only detect limited types of vulnerabilities and struggle to identify unseen ones. 
Among existing tools, DeFiPoser~\cite{arbitrage} adopts the methodology of automatic sketch generation and completion. 
However, its sketches are generated based on limited heuristics, limiting its ability to synthesize anything beyond arbitrage scenarios.

Overall, due to the lack in~\textit{effective searching strategies for attack generation} and~\textit{domain-specific attack validation mechanism}, existing tools cannot effectively synthesize complicated DeFi attacks.  

\begin{figure}[t]
    \centering
    \includegraphics[width=\columnwidth]{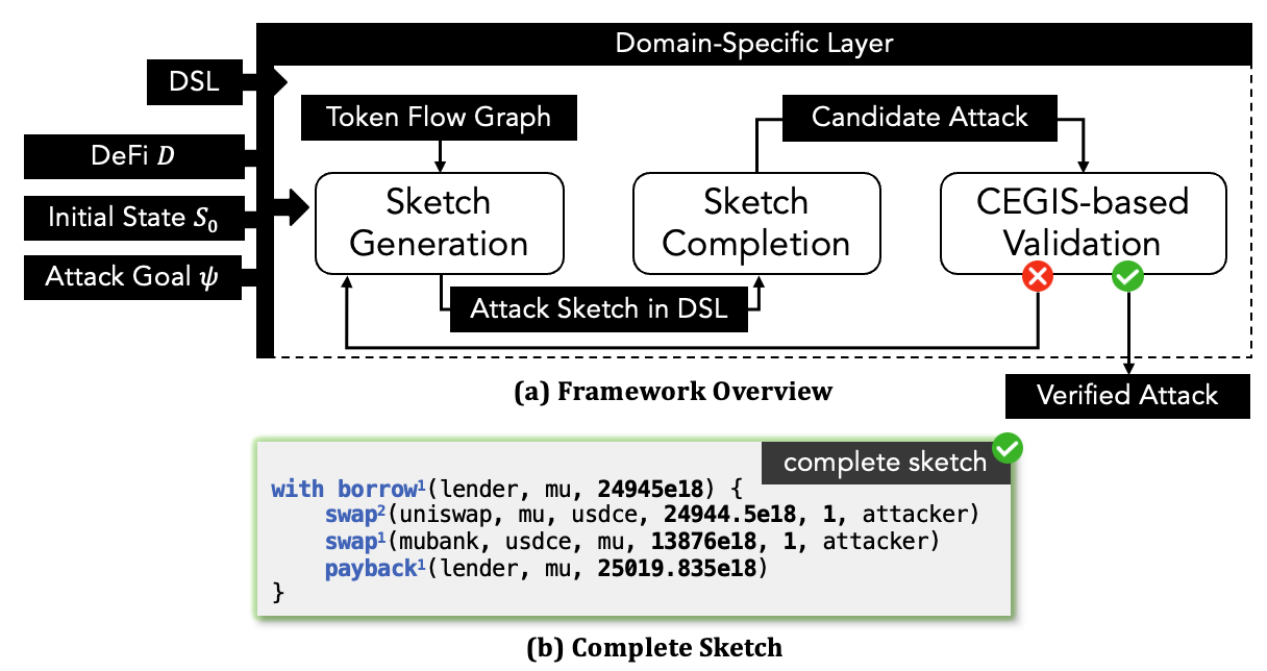}
    \caption{Overview of \tool with the demonstrated completed sketch for the example in \cref{fig:motivating}. 
    In the sketch, \code{swap$^1$} refers to the \code{mn_bond} function. \code{swap$^2$} is achieved through the \code{uniswap.swap} function.  
    }
    \label{fig:overview-cegis}
    \vspace{-5mm}
\end{figure}

\section{Overview of \tool}
\label{sec:overview}

To mitigate the limitations of existing solutions, we design and develop \tool, a novel DeFi-specific attack synthesis technique to uncover various deep logical vulnerabilities in DeFi applications. 
At a high level, \tool follows the attack sketch generation and completion methodology but includes multiple customized designs to enable more effective sketch search and verification.
As shown in \cref{fig:overview-cegis}, we design a domain-specific language to lift the low-level smart contracts into their high-level financial semantics and models (e.g., exchanges, lenders, loans).
Based on our DSL, we first compile DeFi protocols into abstract representations (token flow graph construction), which filter out low-level semantics and constrain the attack sketch space. 
We design an efficient sketch generation method based on the graph reachability in the TFG (sketch generation).
Then, we complete a sketch by compiling it into symbolic constraints and replacing the symbolic variables with concrete assignments using an off-the-shelf solver~\cite{z3} (sketch completion).
Finally, we conduct direct syntax transformation to transform the complete sketches into concrete attacks.
Given that the abstraction process may over-simplify blockchain states and concrete smart contract semantics, we conduct an additional validation step to actually run the synthesized attack.
If an attack cannot satisfy the attack goal, our CEGIS loop will add additional constraints corresponding to the root causes to the solver and avoid similar mistakes in future iterations.

\begin{figure}[!t]
    \includegraphics[width=\linewidth]{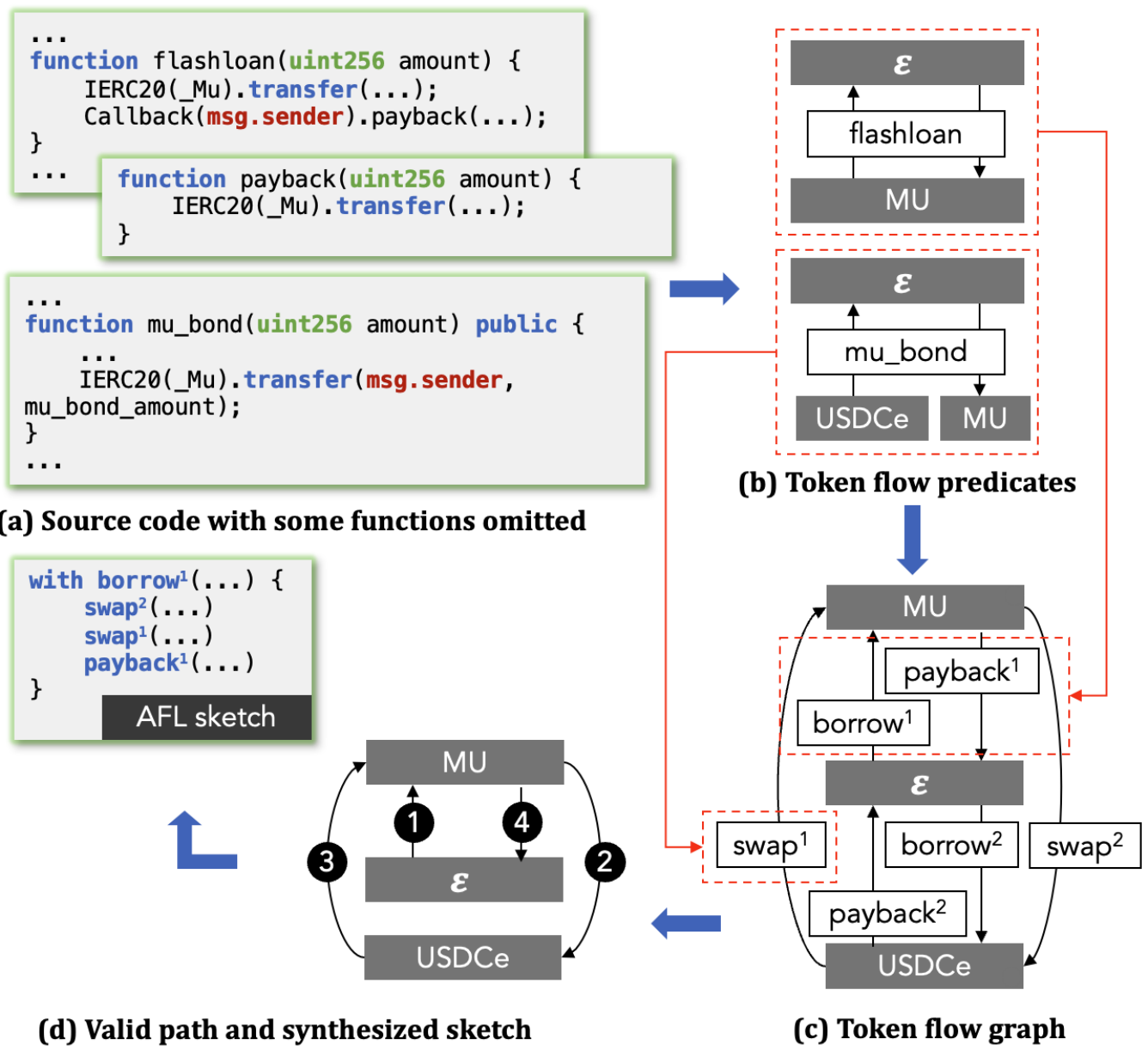}
    \caption{Demonstration of token flow graph construction, graph reachability, and valid attack sketch of \examplename in \cref{fig:motivating}. 
    In the TFG, the token nodes (except $\void$) represent tokens owned by the attacker and edges are financial operators (constructs in abstract financial language).}
    \label{fig:tech-overview}
    \vspace{-3mm}
\end{figure}

\smartparagraph{Token Flow Graph construction (\cref{sec:tfg}).}
The insight of this component is to lift the low-level semantics of smart contracts to their high-level financial models. 
This process filters out a significant portion of solidity semantics, reducing the synthesis space and simplifying the validation process.
To do so, we first define \emph{Abstract Financial Language}, a domain-specific language for describing high-level \emph{financial operations} commonly used by DeFis such as swap, borrow, payback, transfer, etc. 
Then given a DeFi protocol, \tool \emph{lifts} it to a \emph{Token Flow Graph} (TFG).
As we will show later, this TFG helps develop effective strategies for attack sketch synthesis. 
Motivating by prior work~\cite{jungloid,morpheus,tygar} in type-directed program synthesis, we design each node to represent a certain type of token in DeFi.
To avoid and simplify the complexity due to multi-party communication, we also introduce the $\void$ token, a special node that represents tokens from parties other than the current attacker. 
Each edge refers to an operation in our abstract financial language and its source and target nodes represent the tokens that the operation needs to consume and produce, respectively. 
\cref{fig:tech-overview} shows the TFG of the \examplename protocol.
Here the nodes are \exampletoken, \examplestable, and $\void$ (i.e., lender of flash loan).
The edges are possible operations invoking the three smart contracts in \examplename. 
For example, the edge \code{borrow$^1$} from $\void$ to \exampletoken represents one functionality in \code{flashloan} function, which enables borrowing a certain amount of \exampletoken tokens from the lender, i.e., \code{DeFiLender}.

\smartparagraph{Sketch generation (\cref{subsec:6.2}).}

Given a TFG of a victim protocol, an attack goal $\psi$ and an initial state $S_0$ are both expressed as first-order logic constraints, with $S_0$ being satisfied by the initial blockchain state and $\psi$ being expected to be satisfied after the attack program's execution.
The goal of this step is to synthesize an incomplete program $P$ in abstract financial language such that $P(S_0) \models \psi$.

Intuitively, an attack sketch $P$ outlines the key financial steps to achieve the attack goal $\psi$. 
Given the huge space, we need to develop an effective search strategy that only enumerates the sketches that are likely to be successful. 
To do so, we model the problem of achieving the attack goal as a readability problem in our TFG.
We then design a customized graph readability algorithm to efficiently enumerate candidate sketches that conform with the attack goal.


In our motivating example, the attack goal is:
\begin{equation}
    \label{eq:attackgoal}
    B^{\mathit{usdce}}_{t_2} - B^{\mathit{usdce}}_{t_1} > 0,
\end{equation}
stating states the attacker's balance of \examplestable at the end of the execution ($t_2$) should be greater than his initial balance ($t_1$). The details of how to infer the attack goal will be introduced in \cref{sec:impl}.

The attack sketch shown in~\cref{fig:overview-cegis} is a feasible candidate sketch by taking the reachable path of $\void \rightarrow \exampletoken \rightarrow \examplestable \rightarrow \exampletoken \rightarrow \void$ in the TFG.

\smartparagraph{Sketch completion (\cref{subsec:6.3}).}
After synthesizing feasible attack sketches, our next step is to complete the feasible attack sketches by substituting all symbolic variables with concrete assignments with constants or storage variables. 
At a high level, we first design a domain-specific symbolic compilation procedure (motivated by existing solutions~\cite{asplos22-hecate,pldi23-medea,10.1145/2980930.2907962}) that soundly compiles a candidate sketch into a set of constraints that represent the space of all possible concrete attacks. 
Then, we conduct the completion by solving the constraints using an off-the-shelf solver~\cite{z3}. 
The first challenge in this procedure is to constrain the complexity of symbolic constraints such that they are feasible for existing solvers.
As mentioned above, our abstract financial language and token flow graph are proposed for tackling this challenge. 
Representing the victim protocol and attack sketches in our abstract financial language significantly simplifies the constraints.  
The second challenge is how to leverage cases that fail to pass the verification. 
We tackle this by integrating a CEGIS (Counter Example-Guided Inductive Synthesis) loop into the synthesis process.
This step first conducts direct syntax transformation to map the synthesized attack from our abstract financial language back to solidity code.
It then deploys and executes the attack code using foundry framework~\cite{foundry} to test whether the attack goal is achieved in a simulated environment.
It constructs a knowledge base and iteratively adds the root causes of the failed attempts.
We will transform root causes as additional constraints to avoid failed sketches in future attempts. 
\cref{fig:overview-cegis} demonstrates a complete attack sketch given by a constrained solver, where the symbolic variables are filled with concrete values.

As demonstrated \cref{fig:overview-cegis}, \tool also requires inputs $\psi$ and $S_0$ written in first-order logic and a final transformation and validation component (See \cref{sec:attack} for more details of these two parts). 

\section{Token Flow Graph}
\label{sec:tfg}

In this section, we present a new graph abstraction for modeling flows of tokens within a DeFi environment, which is used to summarize common DeFi behavior, as well as searching for potential program sketches that satisfy a given attack goal.

\begin{figure}[!t]
    \centering
    \begin{minipage}[t]{\columnwidth}
        \small
        \setlength{\grammarindent}{1in}
        \setlength{\grammarparsep}{0.1cm}
        \begin{grammar}
            <prog> ::= <stmt>+
        
            <stmt> ::= <transfer> | <burn> | <mint> | <swap> | <borrow>

            <transfer> ::= transfer(token: <token>, from: <addr>, to: <addr>, amt: <expr>)

            <burn> ::= burn(token: <token>, from: <addr>, amt: <expr>)

            <mint> ::= mint(token: <token>, from: <addr>, amt: <expr>)

            <swap> ::= swap(market: <addr>, src: <token>, tgt: <token>, in: <expr>, minout: <expr>, to: <addr>)

            <borrow> ::= with borrow(lender: <addr>, token: <token>, amt: <expr>) \{<stmt>+ <payback>\}

            <payback> ::= payback(lender: <addr>, token: <token>, amt: <expr>)

            <balance> ::= balance(token: <token>, of: <addr>)

            <expr> ::= <const> | <op>(<expr>+) | <balance>
            
        \end{grammar}
        \[
            \begin{array}{c}
                \langle const \rangle \in \textbf{constants} \quad \langle op \rangle \in \textbf{operators} \\ \langle addr \rangle \in \textbf{addresses} \quad \langle token \rangle \in \textbf{tokens}
            \end{array}
        \]
    \end{minipage}
    \caption{Syntax for our abstract financial language.}
    \vspace{-5mm}
    \label{fig:afl}
\end{figure}

\subsection{Abstract Financial Language (AFL)}
\label{subsec:5.1}

As shown in \cref{fig:afl}, Abstract Financial Language (AFL) is a domain-specific language that is designed to model token flows of common financial operations achieved by DeFi protocols. 
A program $\tuple{\it prog}$ written in AFL corresponds to a sequence of statements composed by the following commonly used financial operators:

\begin{itemize}
    \item $\tuple{\it transfer}$ models a single transfer of a specific amount of a token from one address to another.

    \item $\tuple{\it burn}$ models the destruction of a certain amount of a token from an address.

    \item $\tuple{\it mint}$ models the generation of a certain amount of a token from an address.

    \item $\tuple{\it swap}$ models the exchange of a certain amount of one token to another for an address.

    \item $\tuple{\it borrow}$ models a temporary transfer behavior of a certain amount of a token from a lender to a borrower's address. A $\tuple{\it payback}$ statement should always be paired at the end to model the return of the borrowed tokens.
\end{itemize}

Note that $\tuple{\textit{burn}}$ and $\tuple{\textit{mint}}$ functions are implemented to control the total token supply and liquidity, aiming to stabilize its price. 
These operations are restricted to specific authorized users. 
However, attackers may also leverage these functions via exploitation.

AFL also provides easy syntax and interface for accessing different entities from a DeFi environment, including:
\begin{itemize}[leftmargin=*]
    \item $\tuple{\it addr}$ for referring to one of all available addresses in a given DeFi environment.

    \item $\tuple{\it token}$ for referring to one of all available types of tokens in a given DeFi environment.

    \item $\tuple{\it balance}$ accesses a token's balance in a given address.
\end{itemize}

Note that AFL can represent both benign and malicious behaviors.
We mainly use it to model attackers in this work.
\begin{example}[AFL attack program]
As shown in~\cref{fig:tech-overview}(d), an AFL program may include $\tuple{\it borrow}$ and $\tuple{\it payback}$, interspersed with several $\tuple{\it swap}$ operators in the context. 
It represents the following attack behavior: initially, borrowing MU tokens from another party $\epsilon$, then exchanging MU tokens for USDCe tokens, subsequently swapping these back via another exchange contract, and finally, repaying the borrowed MU tokens to the environment $\epsilon$.
\end{example}

\subsection{Definition of Token Flow Graph}
\label{subsec:5.2}

We propose a Token Flow Graph (TFG) to model changes in amounts of abstract tokens owned by the attacker when interacting with public functions of DeFi protocols. 
It helps filter out low-level semantics of smart contracts and guides the synthesis of attack sketches.
To formally define TFG, we first introduce the following domains:

\begin{itemize}[leftmargin=*]
    \item $\fundom$ is a set of public DeFi functions accessible by the attacker. We assume all non-public functions are resolved by inlining.
    \item $\afl$ contains all AFL operators, e.g., \code{borrow}.
    \item $\token$ is a set of different tokens appearing in a given DeFi protocol, i.e., nodes in TFG.
    \item $\edgedom$ is a set of edges in TFG.
    \item $\constraint$ is a set of behavioral constraints about logical relations between tokens, addresses, and AFL operators.
\end{itemize}

Given the above domains, we define a token flow graph as a tuple $\tfg(\token, \afl, \edgedom, \constraint)$. 
In particular, $\edgedom \subseteq \times \token \times \token \times \afl \times \Phi$ is a set of edges connecting tokens, where each edge is associated with an AFL operator. 
For clarity in presentation, edges are attached with superscripts, denoting different functions that they are inferred from.

\smartparagraph{Special node $\void$.} 
Intuitively, the nodes of a token flow graph represent assets of the user currently interacting with the DeFi. 
To reflect and simplify the interactions of other participants (e.g., contract owners, other users), each token flow graph has a built-in node $\void \in \token$ that represents tokens of all participants other than the one of interest (i.e., attacker in our problem). 
Such tokens are not directly related to the attacker's goal but are necessary for the construction of an attack.

\begin{example}[TFG for an attacker]
~\cref{fig:tech-overview}(c) depicts a TFG of the \examplename protocol.
For example, an edge labeled with \code{swap$^1$} indicates that the attacker could exchange \examplestable for \exampletoken through the function \code{mu_bond} in~\cref{fig:motivating}.
\end{example}

\subsection{Construction of Token Flow Graph}
\label{subsec:5.3}

Given a DeFi protocol, the key to constructing a token flow graph for one specific user is to generate edges among tokens that the user holds or wants to acquire. 
\tool employs an edge discovery procedure based on program analysis.
It has two steps, first, we define flow predicate and influence rules for generating flow predicates from concrete programs of a DeFi protocol.
Then, we generate edges from the predicates using edge inference rules.
Each generated edge comes with a semantically equivalent AFL operation with its corresponding constraints. 
As illustrated in~\cref{fig:tech-overview}, we first identify the flow predicates in the \code{flashloan} and \code{mu_bank} function, represented as an initial graph.
Then, we apply the edge inference rules to generate the TFG from flow predicates.
For example, the \code{swap$^1$} is deduced from two token flows in \code{mu_bank}. 
Meanwhile, the \code{borrow$^1$} and \code{payback$^1$} are inferred from the \code{flashloan} function. 
To avoid the confusion between AFL statements and actual (solidity) program statements, we use ``operator'' to represent AFL statements $p \in \mathbb{P}$ and ``statement'' to represent actual program statements $s$.  
In what follows, we elaborate on the procedure for flow predicate and edge construction.

\smartparagraph{Flow predicate.} 
denoted by
${\sf flow}(u,x,a,b)$,
indicates $x$ amount of token $u$ flows from address $a$ to address $b$.
A flow predicate serves as a basic building block of AFL operators.
\cref{fig:inf-flow} shows the rules for generating flow predicates from actual (solidity) programs. 
First, we define a \emph{flow state} $\mathbb{W}$ that contains a collection of $s_i:w_i$ pairs where each pair $s_i:w_i$ represents a statement $s_i$ together with its flow predicate $w_i$. 
Note that $\mathbb{W}$ is different from the blockchain state $S$.
For each public function $f \in \fundom$, the \irule{func} rule processes its statements sequentially by performing a sequence of \emph{flow state transitions}. 
Specifically, given the original state $\mathbb{W}$ and a statement $s$, we model the state transition via
$\mathbb{W} \overset{s}{\rightsquigarrow} \mathbb{W}'$,
which indicates that the analysis of statement $s$ results in a new version $\mathbb{W}'$  by adding the flow predicate corresponding to $s$ to $\mathbb{W}$.
Similar to classical symbolic executions~\cite{oyente,mythril,summarybased}, all loops are bounded and unrolled to their corresponding branch statements. 
The \irule{branch} rule then merges updates of $\mathbb{W}$ from both branches. 
Other rules that update $\mathbb{W}$ are: \irule{flow-from}, \irule{flow-to}, \irule{flow-mint} and \irule{flow-burn}, which correspond to public functions in standard interfaces (e.g., ERC20):

\begin{itemize}[leftmargin=*]
\begin{sloppypar}

    \item The \irule{flow-from} rule can be triggered by invocations of ERC20's \code{transferFrom} (or other similar) interface, e.g., \code{IERC20(u).transferFrom(a,b,x)}, which transfers $x$ amount of token $u$ from address $a$ to address $b$.

    \item The \irule{flow-to} rule can be triggered by invocations of ERC20's \code{transfer} (or other similar) interface, e.g., \code{IERC20(u).transfer(b,x)}, which transfers $x$ amount of token $u$ from the current caller (i.e., the address pointed by \code{this} keyword) to address $b$.

    \item The \irule{flow-mint} rule matches invocations of ERC20's \code{mint} (or other similar) interface, e.g., \code{IERC20(u).mint(a,x)}, which produces $x$ amount of $u$ token for address $a$. 
    
    \item The \irule{flow-burn} rule matches invocations of ERC20's \code{burn}  (or other similar) interface, e.g., \code{IERC20(u).burn(a,x)}, which destroys $x$ amount of $u$ token from address $a$. 
    
\end{sloppypar}
\end{itemize}

After parsing the programs of a DeFi protocol with rules in \cref{fig:inf-flow}, we get a set of flow predicates that summarize critical financial behaviors within that protocol. 
\tool then constructs the token flow graph on top of these predicates.

\begin{figure}
    \centering
    \footnotesize
    \begin{mathpar}

        \inferrule{
            f \in \fundom \\ f \equiv s_0; ...; s_n \\ \mathbb{W}_0 \overset{s_0}{\rightsquigarrow} \mathbb{W}_1 \\ ... \\ \mathbb{W}_n \overset{s_n}{\rightsquigarrow} \mathbb{W}_{n+1}
        }{
            \mathbb{W}_0 \overset{f}{\rightsquigarrow} \mathbb{W}_{n+1}
        }{\ \ \sf\bf (func)}

        \inferrule{
             s \equiv  {\sf if}\ \_\ {\sf then}\ f_0\ {\sf else}\ f_1 \\ \mathbb{W} \overset{f_0}{\rightsquigarrow} \mathbb{W}_0 \\ \mathbb{W} \overset{f_1}{\rightsquigarrow} \mathbb{W}_1
        }{
            \mathbb{W} \overset{s}{\rightsquigarrow} \mathbb{W}_0 \cup \mathbb{W}_1
        }{\ \ \sf\bf (branch)}

        \inferrule{
            s \equiv u.{\sf transferFrom}(a, b, x) \\ w \equiv {\sf flow}(u, x, a, b)
        }{
            \mathbb{W} \overset{s}{\rightsquigarrow} \mathbb{W} \cup \{s:w\}
        }{\ \ \sf\bf (flow{\text -}from)}

        \inferrule{
            s \equiv u.{\sf transfer}(b, x) \\ a = {\sf this} \\ w \equiv {\sf flow}(u, x, a, b)
        }{
            \mathbb{W} \overset{s}{\rightsquigarrow} \mathbb{W} \cup \{s:w\}
        }{\ \ \sf\bf (flow{\text -}to)}

        \inferrule{
            s \equiv u.{\sf mint}(a, x) \\ w \equiv {\sf flow}(u, x, \bullet, a)
        }{
            \mathbb{W} \overset{s}{\rightsquigarrow} \mathbb{W} \cup \{s:w\}
        }{\ \ \sf\bf (flow{\text -}mint)}

        \inferrule{
            s \equiv u.{\sf burn}(a, x) \\ w \equiv {\sf flow}(u, x, a, \bullet)
        }{
            \mathbb{W} \overset{s}{\rightsquigarrow} \mathbb{W} \cup \{s:w\}
        }{\ \ \sf\bf (flow{\text -}burn)}

    \end{mathpar}
    \vspace{-3mm}
    \caption{Flow predicates inference rules. $\bullet$ indicates a special address. Note that mint and burn has an implicit constraint that $a$ must belong to a set of authorized addresses.}
    \label{fig:inf-flow}
    \vspace{-5mm}
\end{figure}

\begin{figure}
    \centering
    \footnotesize
    \begin{mathpar}
        
        \inferrule{
            f \equiv ...; s_1; s_2; ... \\\\ s_1 : {\sf flow}(u, x, a, b) \in \mathbb{W} \\ s_2 : {\sf flow}(v, y, b, a) \in \mathbb{W} \\\\ \Phi \equiv u[a] \geq x \land u'[a] \leq u[a] \land v'[a] \geq y \land v'[a] \geq v[a]
        }{
            {\sf edge}(u, v, {\sf swap}, \Phi)
        }{\ \ \sf\bf (edge{\text -}swap)}
        
        \inferrule{
            f \equiv ...; s_1; ...; s_2; ... \\ s_3 \in g \\ {\sf callback}(s_2, g) \\\\ s_1 : {\sf flow}(u, x, a, b) \in \mathbb{W} \\ s_3: {\sf flow}(u, y, b, a) \in \mathbb{W}
        }{
            {\sf loan}(s_1, s_2, s_3)
        }{\ \ \sf\bf (loan)}
        
        \inferrule{
            {\sf loan}(s, \_, \_) \\ s: {\sf flow}(u, x, b, a) \in \mathbb{W} \\\\ \Phi \equiv u'[a] \geq x \land u'[a] \geq u[a]
        }{
            {\sf edge}(\epsilon, u, {\sf borrow}, \Phi)
        }{\ \ \sf\bf (edge{\text -}borrow)}

        \inferrule{
            {\sf loan}(\_, \_, s) \\ s: {\sf flow}(u, x, a, b) \in \mathbb{W} \\\\ \Phi \equiv u[a] \geq x \land u'[a] \leq u[a]
        }{
            {\sf edge}(u, \epsilon, {\sf payback}, \Phi)
        }{\ \ \sf\bf (edge{\text -}payback)}

                \inferrule{
            s : {\sf flow}(u, x, \bullet, a) \in \mathbb{W} \\ \Phi \equiv u'[a] \geq x \land u'[a] \geq u[a]
        }{
            {\sf edge}(\epsilon, u, {\sf mint}, \Phi)
        }{\ \ \sf\bf (edge{\text -}mint)}

        \inferrule{
            s : {\sf flow}(u, x, a, \bullet) \in \mathbb{W} \\ \Phi \equiv u[a] \geq x \land u'[a] \leq u[a]
        }{
            {\sf edge}(u, \epsilon, {\sf burn}, \Phi)
        }{\ \ \sf\bf (edge{\text -}burn)}

        \inferrule{
            s : {\sf flow}(u, x, a, b) \in \mathbb{W} \\ \Phi \equiv u[a] \geq x \land u'[a] \leq u[a]
        }{
            {\sf edge}(u, \epsilon, {\sf transfer}, \Phi)
        }{\ \ \sf\bf (edge{\text -}transfer)}
        
    \end{mathpar}
    \vspace{-3mm}
    \caption{Edge inference rules. We omit the constraint for $b$ in \irule{edge-swap}, \irule{edge-borrow}, \irule{edge-payback}, and $a$, $b$ in \irule{loan}.}
    \label{fig:inf-edge}
    \vspace{-4mm}
\end{figure}

\smartparagraph{Edge construction.}
\cref{fig:inf-edge} shows the rules for constructing edges in a token flow graph.
Recall that the nodes in a TFG are the tokens that the user holds or wants to acquire, as well as the $void$ node, representing all other parties. 
The underlying mechanism of the edge construction procedure is to identify semantic patterns of flow predicates for each AFL construct. 
An edge is represented by
${\sf edge}(u, v, p, \Phi),$
where $u$ and $v$ are addresses, $p \in \mathbb{P}$ corresponds to an AFL operator and $\Phi$ is a set of $p$'s behavioral constraints. 
We have six types of edges corresponding to different financial operators in \cref{fig:afl}. 
We elaborate on their inference rules as follows:

\begin{itemize}[leftmargin=*]
    
    \item The user could exchange tokens with DeFi functions or third-party APIs from Uniswap, decentralized exchanges, etc.
    The \irule{edge-swap} rule captures such a pattern by looking for a pair of consecutive {\em back-and-forth} flows between two addresses. 
    When a \code{swap} edge is fired, e.g.,
        ${\sf edge}(u, v, {\sf swap}, \Phi)$,
    $u$ tokens are sent in exchange for $v$ tokens. We describe such change of tokens for address $a$ using constraints stored in $\Phi$: $\Phi \equiv u[a] \geq x \land u'[a] \leq u[a] \land$, $v'[a] \geq y \land v'[a] \leq v[a]$,
    where $u[a]$ and $v[a]$ denote $a$'s balances of token $u$ and $v$ respectively, while $u'[a]$ and $v'[a]$ denote corresponding balances after firing the edge. 
    This indicates that $a$ needs at least $x$ amount of $u$ token before swapping, and will get at least $y$ amount of $v$ token after. The invocation of such an operation increases $a$'s balance of token $v$ but decreases its balance of token $u$.

    \item As mentioned in \cref{sec:background}, many DeFis provide {\em flash loans}, a unique feature that enables a (malicious or benign) user to borrow tokens without collateral, as long as the user pays back the loan and its interest within one single transaction. To understand the \irule{edge-borrow} and \irule{edge-payback} rules, we first introduce an auxiliary predicate ${\sf loan}(s_1, s_2, s_3)$ for identifying flash loan patterns in DeFi. In particular, the \irule{loan} rule first looks for a statement $s_1$ together with its corresponding flow. Following $s_1$, a \code{callback} statement $s_2$ is then invoked to register a callback function $g$, which allows the borrower to execute dedicated business logic and produce another flow (from statement $s_3$) that pays the original loan. Once a loan pattern is established, the \irule{edge-borrow} and \irule{edge-payback} will be triggered simultaneously and generate corresponding \code{borrow} and \code{payback} edges. As tokens borrowed could come from different sources, we model the type of token to borrow from and return to using the special node $\void$.

    \item Flows of tokens from the special address $\bullet$ are directly translated into \code{mint} edges via the \irule{edge-mint} rule. The edge goes from $\void$ to $u$ token with constraints ensuring sufficient $u$ tokens after the call. Similarly, flows of tokens to the special address $\bullet$ directly construct \code{burn} edges via the \irule{edge-burn} rule.

    \item Other flows that do not fall into the above categories will generate \code{transfer} edges via the \irule{edge-transfer} rule. Specifically, give a flow predicate ${\sf flow}(u,x,a,b)$, the rule generates a token flow edge (from token $u$ to other participants' token clustered in $\void$) labeled with the \code{transfer} operator. The constraint on the edge asserts that \ding{192} the sender should have sufficient tokens and \ding{193} the sender's remaining $u$ tokens decrease after the call. 
    
\end{itemize}

\section{Attack Synthesis}
\label{sec:attack}

Like prior sketch-based synthesizers~\cite{sketch,neo,rosette}, \tool synthesizes candidate attacks through sketch generation and completion.
The core insight behind \tool's synthesis algorithm is two-folded.
The search space of sketch generation is constrained by graph reachability over a DeFi's TFG (\cref{subsec:6.2}), and the state explosion problem in sketch completion is mitigated by our domain-specific compilation rules over AFL's properties (\cref{subsec:6.3}). 
In what follows, we first give an overview of \tool's synthesis algorithm (\cref{subsec:6.1}), followed by our attack sketch generation (\cref{subsec:6.2}) and sketch completion (\cref{subsec:6.3}) algorithms.

\begin{algorithm}[t]
    \caption{Attack Synthesis}
    \begin{algorithmic}[1]
    \small 
        \Procedure{\textsc{AtkSyn}}{$D, S_0, \psi$}
            \State \textbf{Input:} DeFi $D$, Initial State $S_0$, Attack Goal $\psi$
            \State \textbf{Output:} Attack Program $P$ or $\bot$
            \State $\kappa \gets \top$ \Comment{initialize knowledge base}
            \State $\tfg \gets \textsc{GraphGen}(D, S_0)$ \Comment{construct token flow graph}
            \While{$\tilde{P} \gets \textsc{SketchGen}(S_0, \psi, \tfg, \kappa)$} \Comment{enumerate AFL sketch}
                \State $\delta \gets \textsc{CnstGen}(\phi, R)$ \Comment{generate constraints from sketch}
                \While {$\mu \gets {\sf solve}(S_0 \land \psi \land \kappa \land \delta))$} \Comment{get model}
                    \If {$P \gets {\sf complete}(S_0, \tilde{P}, \mu)$} \Comment{attack instantiation}
                        \If {$P(S_0) \models \psi$} \Comment{validate attack program $P$}
                            \State \Return $P$
                        \Else
                            \State $\kappa \gets \kappa \land \lnot {\sf muc}(P(S_0) \models \psi)$ \Comment{update KB}
                        \EndIf
                    \EndIf
                \EndWhile
            \EndWhile
            \State \Return $\bot$
        \EndProcedure
    \end{algorithmic}
    \label{alg:attack-synthesis}
\end{algorithm}

\subsection{Overview of the Synthesis Algorithm}
\label{subsec:6.1}

\cref{alg:attack-synthesis} shows \tool's top-level attack synthesis algorithm. 
Given a DeFi protocol, its initial state, and an attack goal (in first-order logic), the synthesis algorithm incorporates a two-phased loop, where phase one (line 6) enumerates attack sketches and phase two (line 8) completes concrete attack programs.

\begin{figure}[!t]
    \centering
    \begin{align*}
        \psi\quad ::=\quad& e\ \mid\ \lnot \psi\ \mid\ \psi \land \psi \\
        e\quad ::=\quad& p(\vec{x}, \vec{c})\ \mid\ e_1 \diamond e_2 \ \mid e_1 \odot e_2 \\
    \end{align*}
    \vspace{-3em}
    \begin{align*}
        x \in \textbf{variables}\quad c \in \textbf{constants}\quad p \in \textbf{predicates} \\
        \quad \diamond \in \{+,-,*\} \quad \odot \in \{=,\ge, <\}
    \end{align*}
    \vspace{-2em}
    \caption{Syntax for attack goal language. $\vec{x}$ and $\vec{c}$ represent none or more parameters.}
    \label{fig:atk-lang}
    \vspace{-4mm}
\end{figure}

\smartparagraph{Initial state and attack goal.}
\cref{fig:atk-lang} shows our specification language for expressing initial states and attack goals. 
Initial states and attack goals are expressed through logical expressions over storage variables $x_i$ or constants $c$ in the DeFi environment, e.g., user balances ($B^{\mathit{usdce}}_{t_2}$), blockchain timestamps, \code{msg.sender} etc. 
A complex logical expression $e$ can be composed by arithmetic and logical operators over atomic expressions and custom predicates.
\tool converts attack goals into their corresponding first-order logic formulas via syntax-directed translation. 
For queries that refer to symbols and quantifiers in the program, \tool uses skolemization to make them quantifier-free or reject them otherwise.

\smartparagraph{The main loops.} 
Using the rules in \cref{fig:inf-edge}, the algorithm first constructs a token flow graph from the given DeFi protocol and initial state (line 5). 
It then invokes an enumeration procedure {\sc SketchGen} (\cref{subsec:6.2}) that iteratively searches for candidate attack sketches $\tilde{P}$ (line 6). 
Each sketch $\tilde{P}$ is then compiled by {\sc CnstGen} into constraints $\delta$ that form SMT queries whose solution corresponds to the choices of missing arguments in the attack sketch (line 7). 
\tool enumerates the solution (a.k.a. {\em model}) of these queries (line 8).
Then, \tool completes the attack sketch $\tilde{P}$ and transforms it into a concrete attack program $P$ through direct syntax transformation (line 9). 
The algorithm then validates the effectiveness of the attack, by executing it from the initial state and checking whether the attack goal is satisfied (line 10). 
It returns the concrete attack program $P$ upon passing the validation; otherwise, it invokes a conflict-driven clause learning (CDCL) call (line 13) and moves to the next available candidate.

\smartparagraph{Conflict-driven learning and knowledge base.}
To avoid past mistakes, the algorithm also incorporates a knowledge base $\kappa$ (line 4) that keeps track of constraint clauses that are responsible for each failed validation (line 13).\footnote{${\sf muc}$ stands for ``minimum unsat core''. This corresponds to the feature of {\em unsat core} computation, which is broadly available in modern SMT solvers.}
Similar to previous works on conflict-driven program synthesis~\cite{neo,cav20-concord}, this allows \tool's synthesis algorithm to avoid previously failed cases (by associating the ``root cause'' with corresponding constructs in a candidate program) and refine them for better candidates. 
As such, the knowledge base $\kappa$ is passed as the argument of sketch generation (line 6).

\begin{algorithm}[t]
    \caption{Attack Sketch Enumeration}
    \begin{algorithmic}[1]
        \small
        \Procedure{SketchGen}{$S_0, \psi, \tfg, \kappa$} 
            \State {\bf Input:} Initial State $S_0$, Attack Goal $\psi$, TFG $\tfg$, Knowledge Base $\kappa$
            \State {\bf Output:} Attack Sketch $\tilde{P}$ or $\bot$
            \State {\bf Assume:} $\tfg = (\token, \afl, \edgedom, \constraint)$

            \State $R \gets \{\}$ \Comment{initialize reachable path as ordered set}
            \State $T, \Omega \gets {\sf init}(\tfg, S_0)$ \Comment{initialize token worklist $T$ and constraint store $\Omega$}

            \While {${\sf\bf choose}\ t \in T$} \Comment{choose and remove a token from $T$}
                \State $E \gets \{ e \mid \forall e \in \edgedom\ .\ e \equiv {\sf edge}(t, *, *, *) \}$ \Comment{neighboring edges}
                \ForEach {$e \in E$}
                    \State {\bf if} ${\sf unsat}(\Omega \land \kappa \land e.\Phi)$ {\bf then} {\bf continue}
                    \State $T \gets T \cup \{ e.{\sf out} \}$ \Comment{include output node to worklist}
                    \State $\Omega \gets \Omega \land e.\Phi$ \Comment{update constraint store}
                    \State $R \gets R \cup \{ e \}$ \Comment{add edge to reachable path}
                    \If {$\alpha(\psi) \subseteq T$} 
                        \State $\tilde{P} \gets (e.{\sf op} \mid \forall e \in R)$ \Comment{convert graph to sketch}
                        \State \Return $\tilde{P}$
                    \EndIf
                \EndFor
            \EndWhile
            
            \State \Return $\bot$
        \EndProcedure
    \end{algorithmic}
    \label{alg:sketch-generation}
\end{algorithm}

\subsection{Attack Sketch Generation via Graph Reachability Analysis}
\label{subsec:6.2}

To generate an attack sketch, \tool performs reachability analysis over the TFG and enumerates a reachable {\em path} that consists of multiple edges in the TFG.
The path points from some initial token node (typically $void$, indicating the attacker does not hold that token) to a target token node that the attacker aims to acquire. 
Here, each edge is attached with an AFL operator $p$ and a behavioral constraint $\Phi$ that encodes the pre- and post-condition of triggering $p$ (\cref{fig:inf-edge}).

\smartparagraph{Goal-directed reachability analysis.}
An attack goal $\psi$ in \cref{fig:atk-lang} specifies a logic formula over account balances with \emph{target token(s)} of interest to the attacker.
To satisfy the goal, a feasible sketch has to end up with states that "produce" the target token(s) in $\psi$, by firing a sequence of AFL operators in a path $R$.
Formally speaking, a feasible sketch corresponds to a path in the token flow graph that satisfies the following conditions:
\begin{enumerate}
    \item Satisfiability condition: whether the behavioral constraints $\Phi$ along the path $R$ can be satisfied, and
    \item Coverage condition: whether the path $R$ covers the target token(s) in the attack goal (denoted by $\alpha(\psi)$).
\end{enumerate}

\smartparagraph{Sketch enumeration.} 
Given a token flow graph along with its initial state, attack goal, and knowledge base, the algorithm returns an attack sketch $\tilde{P}$ corresponding to a reachable path.
It consists of a sequence of AFL operators on tokens defined in the TFG. 
The algorithm's main loop (line 7-16) is based on a worklist mechanism that gradually refines the current path until a reachable one is constructed. 
Initially an empty path $R$, together with the token worklist $T$ and constraint store $\Omega$ is created (line 5-6), where $T$ is initialized as tokens that the attacker holds, and $\Omega$ stores constraints converted from initial state $S_0$.
If the attacker does not hold any tokens in the TFG, we initialize $T$ with $\void$.

At each step of the main loop, a token $t$ is first chosen from the worklist $T$ (line 7). 
Then, for each edge $e$ that starts from $t$ (lines 8-9), the algorithm ensures the satisfiability condition is met by checking the conjunction of three sets of constraints using the Z3 solver (line 10); otherwise, it continues with the next available edge. 
For a satisfiable edge $e$, the algorithm updates the token worklist by adding its output token $e.o$, the constraint store by adding its constraint $e.\phi$ (the constraint of triggering its corresponding operator), and the reachable path set $R$ by adding $e$ (lines 11-13).
Then, it checks for the coverage condition by seeking the existence of target tokens from $R$ (line 14). 
The path $R$ is finally converted into an attack sketch $\tilde{P}$ and return if the coverage condition is met (line 15); otherwise, the algorithm keeps trying for the next pair token $t$ and edge $e$ until it finds a satisfiable one or terminate by exhaustion.
Note that every time a valid sketch $\tilde{P}$ is found and returned, the following lines in~\cref{alg:attack-synthesis} will be invoked.
If $\tilde{P}$ fails to achieve the attack goal, the corresponding root cause will be added to $\kappa$ and fed back to {\sc SketchGen}.
The $R$, $T$, $\Omega$ will be reinitialized for generating a new sketch and $\kappa$ ensures that the algorithm avoids the previously failed sketches.

\begin{example}[Attack sketch generation]
In~\cref{fig:tech-overview}(d), a reachable path on the token flow graph begins at the $\void$ node, representing a common scenario where the attacker initially possesses no tokens and must borrow from other entities (\ding{182}). 
Navigating through the graph (\ding{183} - \ding{184}), the attacker is then required to repay the borrowed tokens to prevent execution failure by ending with calling payback and going back to the start node $\ding{185}$. 
The sequence of corresponding operators (\code{borrow} -> \code{swap} -> \code{payback}) along this generated path constitutes a viable sketch candidate for executing the attack.
\end{example}

\subsection{Sketch Completion via Domain-Specific Compilation}
\label{subsec:6.3}

We aim to compile the sketch into a constraint system whose solution results in the completion
of an attack program. 
In particular, using our AFL semantics, we derive a domain-specific compilation that translates the
invocation of each AFL operator into high-level constraints. 
Our constraints are much easier to
solve as they only track the side effects of AFL operators over the attacker’s account balances and filter out low-level semantics of the original DeFi.

\cref{fig:inf-constraints} shows the inference rules for generating constraints of different AFL operators defined in~\cref{fig:afl}.
The rules derive judgments of the form $p \Downarrow C$, where $C$ corresponds to the set of constraints obtained by \emph{symbolically evaluating} an AFL operator $p$. 
For simplicity, we use two macros ${\uparrow}(u, a, x)$ and ${\downarrow}(u, a, x)$ to denote the constraints for describing a balance increase and decrease of amount $x$ of the token $u$ at address $a$, which compiles to $u'[a] = u[a] + x$ and $u'[a] = u[a] - x$, where $u[a]$ and $u'[a]$ denotes the balance of token $u$ for address $a$ before and after evaluating the corresponding operator $p$.

Each inference rule in~\cref{fig:inf-constraints} models the change of account balances caused by the corresponding AFL operator. 
For instance, the \irule{c-transfer} rule generates constraints to assert the increased and decreased amounts of recipient and sender, respectively. 
The \irule{c-swap} rule states that from a sender's view (address $a$), the balance of its source token will decrease and its target token will increase. 
The recipient's (address $b$) case is the inverse.

In addition to modeling balance changes, the rules also model financial features for certain operators. 
For example, for \code{swap} operator, besides the macro $\varsigma(a,u,v,x,y,b)$ that describes mutual balance changes between address $a$ and $b$,\footnote{This compiles to ${\downarrow}(u,a,x) \land {\uparrow}(u,b,x) \land {\downarrow} (v,b,y) \land {\uparrow}(v,a,y)$.} we introduce $\rho(x,y)$ to model the {\em invariant} between token pairs in modern automated market makers (e.g., $x \cdot y = k$ in Uniswap).

Meanwhile, for tokens that provide flash loans, the constraint of an additional fee is modeled via $\vartheta(x,y)$, where $x$ is the amount of flash loan and $y$ is the amount of repayment, and $y > x$ in most cases means additional interest is charged in \code{payback}.

Such constraints are inferred in a data-driven way via analysis of massive amounts of real-world transaction data.
Since the arguments of an AFL operator may refer to local variables, we leverage off-the-shelf pointer analysis to resolve their actual locations. 

Given a sketch $\tilde{P} = (p_1, p_2, ... )$, the constraints of $\tilde{P}$ are obtained by 
1) applying the inference rule on each $p_i$ and then 2) conjoining all the resulting constraints together:
$\textsc{CnstGen}(S_0, \tilde{P}) = \mathsf{foldl}(S_0, \mathsf{map}(\tilde{P}, \Downarrow ), \land)$.

\begin{figure}[t!]
    \centering
    \footnotesize
    \begin{mathpar}
        \inferrule{
            p \equiv \mathsf{transfer}(u,a,b,x)
        }{
            p \Downarrow {\downarrow}(u, a, x) \land {\uparrow}(u, b, x)
        }{\ \ \sf\bf (c{\text -}transfer)}

        \inferrule{
            p \equiv \mathsf{burn}(u,a,x)
        }{
            p \Downarrow {\downarrow}(u, a, x)
        }{\ \ \sf\bf (c{\text -}burn)}

        \inferrule{
            p \equiv \mathsf{mint}(u,a,x)
        }{
            p \Downarrow {\uparrow}(u, a, x)
        }{\ \ \sf\bf (c{\text -}mint)}

        \inferrule{
            p \equiv \mathsf{swap}(a,u,v,x,y,b)
        }{
            p \Downarrow \varsigma(a, u, v, x, y, b) \land \rho(x, y)
        }{\ \ \sf\bf (c{\text -}swap)}

        \inferrule{
            p \equiv \mathsf{borrow}(u,a,b,x)
        }{
            p \Downarrow {\downarrow}(u, a, x) \land {\uparrow}(u, b, x)
        }{\ \ \sf\bf (c{\text -}borrow)}

        \inferrule{
            p \equiv \mathsf{payback}(u,a,b,y)
        }{
            p \Downarrow {\downarrow}(u, a, y) \land {\uparrow}(u, b, y) \land \vartheta(x,y)
        }{\ \ \sf\bf (c{\text -}payback)}
        
    \end{mathpar}
    \vspace{-3mm}
    \caption{Domain-specific constraint compilation rules.}
    \label{fig:inf-constraints}
    \vspace{-5mm}
\end{figure}

\section{Implementation}
\label{sec:impl}

We have implemented \tool in Python with a back-end constraint solver (Z3~\cite{z3} version 4.12.2). To fetch the concrete state and verify the feasibility of the attack sketches, \tool integrates Foundry~\cite{foundry} to interact with the blockchain. In what follows, we elaborate on various aspects of our implementation.

\smartparagraph{Attack goal generation.} In real-world cases, an attack who seeks for financial gains would try spending no assets when launching an attack (i.e., launching an attack with no cost), which requires appropriate choices of an attack goal. To achieve this, \tool automatically gathers all \textit{stablecoins} involved in the target DeFi protocol from their on-chain storage variables, and includes them as potentially hackable assets into the attack goal. \tool then tries to solve a feasible attack program for each hackable asset.

For example, in the MUMUG protocol mentioned in \cref{sec:motivating}, users could spend USDCe to buy MU without any incentivization. We abbreviate the beginning and ending balance of USDCe of an attacker as \(B^{\mathit{usdce}}_{t_1}\) and \(B^{\mathit{usdce}}_{t_2}\), accordingly.
The contract invariant can then be formalized as:
\begin{equation*}
    B^{\mathit{usdce}}_{t_2} - B^{\mathit{usdce}}_{t_1} \le 0,
\end{equation*}
and the attack goal, formalized as~\eqref{eq:attackgoal}, is to find a concrete exploit that violates the above invariant.


\smartparagraph{Inference of token flow.}
\tool compiles a DeFi protocol (e.g., in Solidity) to its AFL representation via the following steps:
\begin{itemize}[leftmargin=*]

    \item A static analysis procedure (e.g., provided by Slither~\cite{slither}) is invoked to first generate machine-readable intermediate representation (IR), e.g., Slither IR, of the DeFi protocol.

    \item Token flows can then be identified from the generated IR via standard interfaces, e.g., ERC20~\cite{ERC20TokenStandard2015}, ERC721~\cite{ERC721TokenStandard2018}, and ERC1155~\cite{ERC1155MultiTokenStandard2018} in Solidity/EVM, and extracted on statement level, as described by \cref{fig:inf-flow}.

    \item \tool then infers the corresponding AFL functions from the identified token flows via rules defined in \cref{fig:inf-edge}.

    
    
    
\end{itemize}


\section{Evaluation}
\label{sec:eval}


All experiments are conducted on an Amazon EC2\textregistered\ instance with an AMD EPYC 7000\textregistered\ CPU, 8 Cores, and 64G of memory running on Ubuntu 20.04. 
We set the default timeout for the solver as 3 hours.
This number is obtained by observing the performance of \halmos. 
In most cases, it either finishes the process at around 2-3 hours or fails completely. 
Our evaluation plans to answer the following research questions:
\begin{itemize}[leftmargin=*]
    \item (RQ1): How does \tool perform compared to SOTA tools?
    \item (RQ2): Is \tool effective in detecting known vulnerabilities?
    \item (RQ3): How effective are the two key designs of \tool and whether \tool will introduce false positives?
    \item (RQ4): Can \tool be useful in detecting zero-day vulnerabilities?
\end{itemize}

\subsection{Detecting Known Vulnerability (RQ1\&RQ2)}
\label{subsec:7.2}

\smartparagraph{Benchmark.}
To evaluate \tool on known vulnerabilities, we select our benchmarks from the DeFiHackLabs dataset~\cite{DeFiHackLabs}, which keeps track of all DeFi hack incidents in the past. 
The DeFiHackLabs dataset records 389 incidents (at the time of the submission). 
We consider a subset of 200 benchmarks from Jan 2022 to July 2023 and exclude old benchmarks before 2022 because they depend on outdated versions of the Solidity compiler. 
Furthermore, we exclude benchmarks from one of these categories: 
a) closed source, 
b) common vulnerabilities (as referred in \cref{sec:background}) such as integer overflow, reentrancy, access controls, etc., 
and 
c) insider hacks due to losing primary keys or misconfiguration. 
Our dataset ends up with 34 representative benchmarks. 
To get better insights into the root causes of the benchmarks, we also categorize them into four types of logical flaws: 
\ballnumber{1} \emph{Token Burn} (TB), where the attack can indirectly mint or burn the victim's tokens by calling the corresponding mint or burn function through other public functions (similar to privilege escalation); 
\ballnumber{2} \emph{Pump \& Dump} (P\&D): inflating the price of a token through abnormal financial transactions (e.g., spitefully  inflates the token price through substantial purchases);
\ballnumber{3} \emph{Price Discrepancy} (PP), which allows the attack to generate profits based on the price difference of the same token pair in different smart contracts (e.g., \examplename);
\ballnumber{4} \emph{Swap Rate Manipulation} (SR): the attack can directly or indirectly influence the swap rate between multiple token pairs in the same smart contract. 
In total, the selected benchmark vulnerabilities have cost $>\$21$M of losses.

\smartparagraph{Baseline.}
As discussed in \cref{sec:motivating}, there are two possible existing solutions for our problem. 
We select one SOTA tool for each solution as our baseline method.
For sketch generation and completion, we use our sketch generation method (given that no existing tool can strategically generate sketches) and use \halmos~\cite{halmos}, the SOTA symbolic reasoning tool for DeFi, for sketch completion. 
We select \ityfuzz\cite{ityfuzz}, the SOTA tool for cross-contract fuzzing, as our baseline method for the fuzzing solution. 
Note that an existing DeFi security tool, DeFiPoser~\cite{arbitrage}, also follows the sketch generation and completion methodology but can only be applied to arbitrage (PP in our benchmark).
Due to its limited scope and lack of open-source implementation, we do not include it as our comparison baseline. 

We run \tool, \halmos, and \ityfuzz on the selected benchmarks using the same computational resource and timeout limit mentioned above. 
We report the runtime needed for each method to detect each selected vulnerability.
We also report the average run time over the success cases (the vulnerabilities that are detected within the time limit) and the overall success rate to assess the effectiveness and efficiency of each tool.

\newcolumntype{?}{!{\vrule width 1pt}}
\begin{table}[t!]
    \centering
    \caption{Running time of \tool vs. \halmos and \ityfuzz on the selected benchmark. ``TO'' means the tool cannot find a valid attack for the corresponding vulnerability within the time limit, and ``NA'' means the benchmark is not supported.
    }
    \vspace{-4mm}
    \resizebox{1\columnwidth}{!}{
    \begin{tabular}{c?c|c|c|c|c}
        \Xhline{1.0pt}
         Name & Category & \tool & \ityfuzz & \halmos & \# of Sketches \\ \Xhline{1.0pt}
         AES  & TB & 25.1s & 27.0s & TO & 16 \\ \hline 
         BGLD & TB & 25.1s & 172.0s & TO & 16 \\ \hline 
         BIGFI & TB & 25.1s & 511.0s & TO & 16 \\ \hline
         BXH & P\&D & 350.5s & TO & TO & 38 \\ \hline
         Discover & PP & 325.1s & NA & 10251.3s & 16 \\ \hline
         EGD & P\&D & 25.6s & 2.0s & TO & 56 \\ \hline
         MUMUG & PP & 300.2s & NA & 7681.7s & 16 \\ \hline
         NOVO & TB & 25.1s & 81.0s & TO & 16 \\ \hline
         OneRing & P\&D & 25.3s & TO & TO & 32 \\ \hline
         RADTDAO & TB & 25.1s & 627.0s & TO & 16 \\ \hline
         RES & SR & 25.2s & 3.0s & TO & 16 \\ \hline
         SGZ & SR & 25.2s & TO & TO & 30 \\ \hline
         ShadowFi & TB & 25.1s & 1757.0s & TO & 16 \\ \hline
         Zoompro & SR & 25.2s & TO & TO & 36 \\ \hline
         NXUSD & P\&D & TO & TO & TO & -- \\ \hline
         NMB & P\&D & 626.3s & TO & TO & 21 \\ \hline
         Lodestar & P\&D & TO & TO & TO & -- \\ \hline
         SafeMoon & TB & 25.1s & TO & TO & 16 \\ \hline
         Allbridge & PP & TO & NA & TO & -- \\ \hline
         Swapos V2 & SR & 25.7s & 182.3 & 6322.0s & 80 \\ \hline
         Axioma & P\&D & 25.7s & TO & TO & 22 \\ \hline
         0vix & PP & TO & NA & TO & -- \\ \hline
         NeverFall & P\&D & 100.7s & TO & TO & 16 \\ \hline
         SellToken02 & P\&D & 26.0s & TO & TO & 180 \\ \hline
         LW & PP & 25.1s & NA & TO & 16 \\ \hline
         UN & TB & 25.1s & 10.1s & TO & 16 \\ \hline
         CFC & TB & 625.2s & TO & TO & 90 \\ \hline
         Themis & P\&D & TO & TO & TO & -- \\ \hline
         Bamboo & TB & 25.1s & 5.2s & TO & 16 \\ \hline
         LUSD & P\&D & 25.1s & TO & TO & 16 \\ \hline
         RodeoFinance & PP & TO & NA & TO & -- \\ \hline
         Carson & PP & TO & TO & TO & -- \\ \hline
         XAI & TB & 25.1s & TO & TO & 16 \\ \hline
         Hackathon & TB & 25.1s & TO & TO & 16\\ \hline
         & Succ. rate & 79\% (27/34) & 32\% (11/34) & 9\% (3/34) & Avg. \# \\ \hline 
         & Avg. Time & 105.9s & 307.1s & 8085.0s & 31.7 \\ 
         \Xhline{1.0pt}
    \end{tabular}
    }
    \label{tab:comp}
    \vspace{-5mm}
\end{table}

\smartparagraph{Results.} 
\cref{tab:comp} shows the main results of the three tools on the selected 34 benchmarks.
Here, the first two columns represent the name and category of each benchmark. 
Columns 3-5 show the running time of \tool, \halmos, and \ityfuzz, respectively. 
We treat ``TO'' and ``NA'' as failure cases.
\tool successfully synthesizes the attack programs for 79\% benchmarks whereas \halmos and \ityfuzz only solve 9\% and 32\% benchmarks, respectively.
This result demonstrates that by modeling financial logic, \tool is significantly more effective in synthesizing DeFi logical bugs compared to SOTA tools. 
These tools often struggle to capture application logic and rely on brute-force solutions.
Furthermore, \tool is also more efficient than baseline approaches in that it takes an average time of 105.9 seconds to solve 27 benchmarks. 
In comparison, \halmos takes an average time of 8,085.0 seconds to solve three benchmarks and \ityfuzz takes an average time of 307.1 seconds to solve 11 benchmarks from our dataset.
\tool's high efficiency benefits from its strategical sketch generation, which improves the search efficiency, and its domain-specific compilation, which simplifies the constraints.
The "\# of Sketches" column shows the number of sketches generated by \tool. On average, 31.7 sketches are generated, with a maximum of 180 for benchmark SellToken02 and a minimum of 16.

We took a closer look at \tool's performance regarding different categories of benchmarks and realized that \tool performs better on TB and SR than PD and PP. Compared to other types of vulnerabilities, PD and PP usually require a series of repeated arbitrages and flash loans to reach the preset profit in the attack goal. Therefore, the number of parameters and steps in those benchmarks is larger and it takes longer for the synthesizer to enumerate and verify candidate attacks.

\subsection{Ablation Study and False Positive (RQ3)}
\label{subsec:7.2.2}

\smartparagraph{Benefit of domain-specific compilation.}
Given that \halmos and \tool use the same sketch generation procedure, \halmos is equivalent to the ablative version of \tool without domain-specific compilation. In other words, the difference in their performance as shown in \cref{tab:comp} is mainly caused by the different mechanisms of symbolic compilation.
\halmos uses a general-purpose compilation to symbolically evaluate each benchmark using \emph{concrete semantics} of solidity. 
It only solves the three easiest benchmarks.
\halmos generates 1,360 and 2,179 constraints for Discover and MUMUG, whereas \tool only generates 64 and 120 constraints, respectively. 
This confirms that our domain-specific compilation significantly reduces the amount of generated constraints, greatly simplifies the solving process, and thus enables more successful cases.  

\smartparagraph{Benefits of sketch generation.}
To further evaluate the effectiveness of our attack generation algorithm, we replace it with a straightforward breadth-first search and keep all other comments the same.
This method brute forces all operators to have a certain length, starting from a length of one, where each operator is treated as a program sketch. 
Our result shows that \tool times out on all benchmarks. 
This is due to the straightforward solution enumerating a huge number of sketches, causing time out. 
The result verifies the necessity of our sketch generation method in improving the overall efficiency of the synthesis process. 

\smartparagraph{False positives.}
We run \tool on 50 benign DeFi protocols, which contain the 34 benchmarks in \cref{tab:comp} after fixing the bugs and ten popular DeFi protocols from Defillama (Lido~\cite{lido}, MakerDAO~\cite{makerdao}, Aave~\cite{aave}, etc.). 
We treat the 10 popular protocols as benign because they pass commercial auditing.
Our results show that \tool timed out (even after we increased the timeout time to 6 hours) on all those benchmarks and did not find any attacks.
This result validates \tool's capability of avoiding false positives. 

\begin{figure}[!t]
    \includegraphics[width=\linewidth]{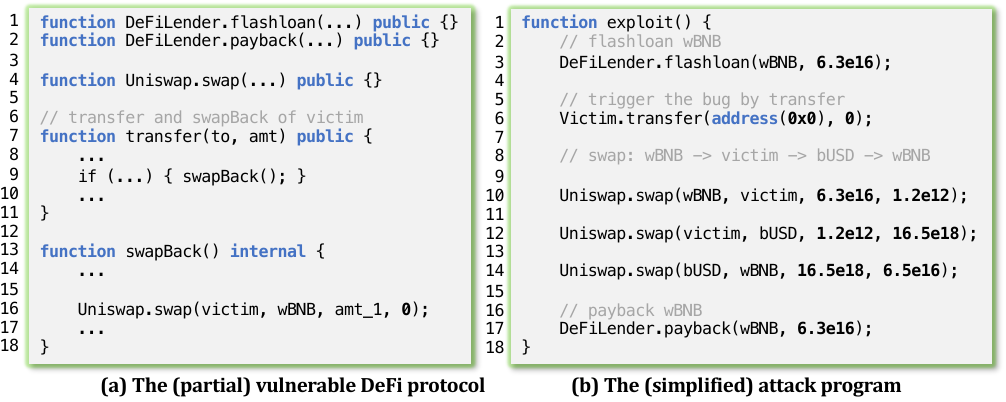}
    \caption{A zero-day vulnerability detected by \tool. ``victim'' stands for the address of the token issued by the Victim contract.}
    \label{fig:zeroday}
      \vspace{-4mm}
\end{figure}

\subsection{Detecting Zero-day Vulnerability (RQ4)}
\label{subsec:7.3}

The BNB chain has gained significant traction recently due to its low transaction fees. However, it also accounts for 30\% of recent exploits, according to professional web3 security reports~\cite{certik-report-2023,merkle-science-2022}. To explore more potential issues, we applied \tool to 5,000 high-profile DeFi protocols on the BNB chain and uncovered 10 previously unknown vulnerabilities, ranging from different types of logical flaws (TB/P\&D/PP/SR). These vulnerable DeFi protocols have a total TVL of 1.1M USD, with the maximum, minimum, and average TVLs being 398K, 2.7K, and 10.7K, respectively. In terms of transactions, these protocols have a total of 1.4M transactions, with the most popular one having 1.3M transactions and the most recently deployed one having only 28 transactions. On average, these protocols have 140K transactions, indicating their activity levels.

This result confirms \tool's capability of discovering diverse unseen vulnerabilities, which are challenging for existing pattern matching-based approaches (e.g., DeFiRanger~\cite{wu2021defiranger} and DeFiTainter~\cite{defitainter}).
Furthermore, the attack synthesized by our tool typically involves more than five transaction actions, which are challenging for general-purpose symbolic execution (e.g., \halmos and DeFiPoser~\cite{arbitrage}) and fuzzing tools (e.g., \ityfuzz). 

All bugs found by \tool are reported to, confirmed, and fixed by corresponding project developers through private channels. We help project developers avoid financial loss via three ways:
\begin{itemize}[leftmargin=*]
    \item Use the administrative functions of the protocol to disable the vulnerable public functions.
    \item Lock the protocol and return assets to users.
    \item Upgrade their smart contracts if possible.
\end{itemize}

Here, we illustrate one major vulnerability belonging to SR to show how \tool synthesizes the exploit.
\cref{fig:zeroday} shows the buggy protocol and its exploit generated by \tool. 
The victim protocol has a logical flaw in its token swap mechanism, i.e., \code{swapBack} function that will cause a price change between victim and wBNB.
Specifically, as shown in~\cref{fig:zeroday}(b), \tool generates a program with six concrete function calls. 
Here is the logic to trigger the vulnerability: 
\ballnumber{1} The attacker takes a flash loan of some WBNB tokens by calling \code{DeFiLender.flashloan}. 
\ballnumber{2} The attacker then calls \code{Victim.transfer} to trigger the \code{swapBack}.
As shown in~\cref{fig:zeroday}(a), the internal function \code{swapBack} swaps a certain amount \code{amt_1} of victim to wBNB, causing a devalue of victim and increasing value of wBNB in the \code{Uniswap} contract.
\ballnumber{3} the attacker leverage the price change to swap more victim with the loaned wBNB.
~\ballnumber{4}~\ballnumber{5} Attacker sequentially swaps Victim to bUSD and bUSD to wBNB. 
Given that the attacker gets more victim than usual cases after \ballnumber{3}
This enables the attacker to get more wBNBs than its original loaned amount. 
\ballnumber{6} Eventually, attacker calls \code{DeFiLender.payback} to pay back the flash loan and keep the extra $0.2e^{16}$ wBNB as the profit.
The exploit program plunderers approximately 11\% of the valuable stablecoins (BUSD) in the liquidity pool as the profit. 
\tool spent 318.4 seconds synthesizing this program while neither \halmos nor \ityfuzz synthesizes a comparable solution within the allotted time frame.

\section{Discussion}
\label{sec:discussion}

\smartparagraph{Generalizability and scalability.}
As illustrated in \cref{sec:eval}, \tool can synthesize attacks for various types of logical bugs that current tools cannot detect. 
However, we acknowledge that there are more types of deep logical bugs that our tool has not yet addressed~\cite{zhou2023sok,zhang2023demystifying}. 
So far, these vulnerabilities have been discovered by highly experienced human auditors. 
By extending our TFG construction and compilation rules, \tool can be generalized to address other vulnerabilities as well.
For example, we can introduce a higher order operator that conducts individual AFL operators multiple times to handle erroneous accounting~\cite{zhang2023demystifying}, which requires accumulating a small computational discrepancy multiple times.
Similarly, \tool can also be generalized to common vulnerabilities although they are not our focus.
Our future work will extend \tool to more types of deep logical vulnerabilities.

\cref{sec:eval} demonstrates that \tool significantly outperforms existing tools in synthesizing complicated logical bugs (e.g., the zero-day bug in \cref{subsec:7.3}).
However, we also notice that \tool still fails to synthesize some ultra-complicated cases (\cref{tab:comp}) due to the limited capability of the SOTA solver. 
In our future work, we will explore hybrid approaches that leverage symbolic execution and fuzzing for sketch completion to improve scalability. 
Note that our sketch generation would still be valuable in that it is challenging for fuzzing to generate valid transaction sequences. 


\smartparagraph{Manual efforts.}
So far \tool still requires certain manual efforts for the generation of the attack goal and initial state specification, as well as additional function mappings.
Here, additional function mappings refer to the auxiliary parameters and extra function calls that must be incorporated when mapping an AFL action back to concrete functions. 
These manual efforts are still way lower than the amount of effort needed to summarize patterns from historical attacks or manual auditing. 
In addition, pattern summarization and matching have limited generalizability. 
Our future works will explore automating these steps, such as leveraging deep learning to generate specifications~\cite{DSM} and data mining to extract additional function mappings~\cite{sm}. 


\smartparagraph{Defense.}
As an offensive defense work, our ultimate goal is to uncover more attacks before they actually happen and provide such attacks to DeFi developers and users so that they can improve their protocol or transaction safety.
\tool's capability of providing exploits makes it easier for developers to analyze the root cause and apply proper defenses.  
In general, we can patch the vulnerable protocol or add run-time assertions.
For example, we can fix the bug in \examplename by upgrading the way of deciding converting price between \exampletoken and \examplestable such that the price is robust against the dramatic changes in their reservations. 

\smartparagraph{DSL design choices and VM compatibility.}
The design of the existing DSL (\cref{fig:afl}) as well as the token flow graph (\cref{sec:tfg}), considers a balance among generality, efficiency, and the amount of domain knowledge incorporated. As we show the flexibility of \tool, in practice, one can always lean towards different design choices (e.g., towards more precise domain knowledge) and adjust the DSL and graph structure accordingly. \tool is instantiated in Solidity in our evaluation, which is a programming language supported by any EVM-compatible VMs. Since our DSL is language-agnostic, with sufficient engineering effort, \tool can be instantiated with other VMs (e.g., MoveVM~\cite{movevm}, SVM~\cite{svm}, etc.) as well. 

\smartparagraph{Extension with data-driven approaches.}
During synthesis, \tool has to make decisions on which DeFi protocols and functions to enumerate. While this is still an open problem, compared to a brute-force enumeration, the key insight of \tool is to leverage the token flow graph and attack goal to avoid enumerating choices doomed to fail. Such a core insight naturally gives a potential future extension that leverages data-driven approaches to explore candidates that maximally align with the application logic. We believe the modularity of \tool's procedures opens up new room for enhancement and integration of data-driven approaches.

\smartparagraph{Complex path conditions and statements.} Flow predicates are used to construct token flow graphs, which are used by sketch enumeration. Since a token flow graph over-approximates the behavior of a Solidity program, most complex path conditions and statements, including modeling of access-controls are abstracted away conservatively during the sketch enumeration phase and the precision loss will be recovered in a goal-driven way during the sketch completion phase via a CEGIS procedure.
\section{Related Work}
\label{sec:related}


\smartparagraph{Smart contract vulnerability analysis.}
Existing tools for detecting and analyzing smart contract vulnerabilities can be categorized into either static analysis~\cite{Madmax,ReentrancyOOPSLA,ReentrancyPOPL} or dynamic analysis~\cite{ityfuzz,SMARTIAN,contractFuzz} approaches.
Static tools conduct static analysis or symbolic execution to detect the common vulnerability (mentioned in \cref{sec:background}) that does not require a deep understanding of a DeFi protocol. 
Notably, Securify~\cite{securify} analyzes a smart contract's bytecode and finds pre-defined patterns in its control flow graph corresponding to certain bug types.
Slither~\cite{slither} (also used in \tool) is the most stable and frequently maintained static analysis framework to analyze smart contracts. 
Notable symbolic execution tools include Manticore~\cite{manticore}, Mythril~\cite{mythril}, Solar~\cite{summarybased}, and \halmos~\cite{halmos} (the SOTA).
As demonstrated in \cref{sec:eval}, without effective sketch generation and domain-specific compilation, solely relying on symbolic execution cannot handle deep logical bugs in DeFi protocols. 
Most dynamic and hybrid analysis tools are designed to be used within one smart contract~\cite{contractFuzz,sFuzz,Harvey,sailfish2022}.
Without an understanding of protocol logic, the fuzzers that support cross-contract fuzzing (e.g., ItyFuzz~\cite{ityfuzz}) cannot maintain their effectiveness in DeFi attack synthesis. 

\smartparagraph{DeFi Security.}
The key challenge for DeFi security lies in the larger size and broader scope beyond individual smart contracts as well as the complicated semantics and logic involved.  
Aside from Zhou et al.~\cite{zhou2023sok} which conducts a comprehensive summary of existing DeFi attacks, existing works in this domain mainly follow the methodology of summarizing patterns from existing attack instances and building attack detection tools via pattern matching. 
Specifically, DeFiRanger~\cite{wu2021defiranger} lifts the low-level smart contract semantics to high-level ones and uses them to summarize and express patterns.
FlashSyn~\cite{chen2022flashsyn} leverages numerical approximation to extract patterns from attack transaction sequences and detect suspicious transactions during run time. 
UnifairTrade~\cite{unfairtrades} identifies fragile swap pair implementations as patterns.
DeFiTainter~\cite{defitainter} conducts taint analysis with taint source and target summarized from standard smart contract API templates.
The capability and scalability of these approaches are constrained by the pattern extraction step.
In fact, the above approach can only detect a certain type of price manipulation vulnerability that leverages \inlinecode{swap} to manipulate token prices (e.g., \examplename).  
More recent tools also extend this methodology to other vulnerabilities.  
For example, DeFiCrisis~\cite{gudgeon2020decentralized} introduces strategies for exploiting DeFi governance mechanisms by arranging funding to gain profits. 
TokenScope~\cite{Tokenscope} is designed to detect any inconsistent and phishing behaviors in token applications. 
The technique that most aligned with \tool is DeFiPoser~\cite{arbitrage}, which proposes two strategies to facilitate the generation of exploit for profit.
The first strategy creates sketches using heuristics and then completes them with an SMT solver, while the second strategy identifies potential trades through a method known as negative cycle arbitrage detection.
Due to the limitation in sketch generation, this tool can only work with arbitrage detection, whereas \tool can be applied to a variety of DeFi protocols, detect different financial flaws, and synthesize complex trading sequences.


\smartparagraph{Attack synthesis and exploit generation.} 
The synthesis of cyber-attacks and the automated generation of exploits have been subjects of significant research interest, aiming to understand and mitigate security vulnerabilities. 
The seminal work, AEG~\cite{avgerinos2014automatic}, used symbolic execution techniques to generate the exploit for the shell program. 
Attack synthesis techniques have been applied to many domains, such as Mayhem~\cite{cha2012unleashing} using concolic execution for Linux Kernel, Intellidroid~\cite{wong2016intellidroid} using dynamic analysis and fuzzing for Android, HeapHopper~\cite{eckert2018heaphopper} using bounded model checking for Memory allocator, AASFSM~\cite{pacheco2022automated} using NLP techniques for TCP and DCCP protocols and etc. 
Symbolic execution is a well-adopted technique to generate a specific exploit, which creates a set of constraints based on the original program and then solves them by delegating SMT solvers. 
Compared with a general symbolic execution technique, \tool first benefits from general financial knowledge to eliminate the search space of synthesis efficiently, then do the domain-specific compilation to generate more lightweight constraints for existing SMT solvers to solve, eventually becoming scalable in the DeFi attack synthesis domain.




\section{Conclusion}
\label{sec:concl}

We present \tool, a highly effective attack synthesis framework against deep logical bugs in DeFi protocols.
Different from existing tools that only detect common vulnerabilities in individual smart contracts, \tool effectively models the financial logic in DiFi protocols and synthesizes exploits against logical flows accordingly.   
Our evaluation of 34 benchmark DeFi security attacks demonstrates the advantage of \tool over existing smart contract bug-haunting approaches.
We further show that \tool can uncover ten zero-day vulnerabilities from the BNB chain.
Finally, we demonstrate the effectiveness of \tool's two key designs (sketch generation and completion) and its capability of avoiding false positives. 
From extensive evaluation, we can safely conclude that with domain-specific modeling and compilation, symbolic reasoning can be an effective approach for exploit synthesis against deep logical bugs in DeFi protocols. 

\section{Acknowledgements}
We are truly grateful for the time and effort that the anonymous reviewers invested in reviewing our work and offering valuable feedback.
This work is supported in part by Google Faculty Research Award, Ethereum Foundation Academic Award, NSF 1908494, and DARPA N66001-22-2-4037.
The views and conclusions contained in this document are those of the authors. They should not be interpreted as representing the official policies, expressed or implied, of the funding agencies.


\bibliographystyle{ACM-Reference-Format}
\bibliography{main}

\clearpage
\appendix
\onecolumn

\section{Glossary}
\label{apdx:glossary}

\begin{table}[H]
\centering
\caption{\centering Summary of Notations}
\resizebox{\textwidth}{!}{%
\begin{tabular}{c?l|l}
\noalign{\hrule height 1pt}
{\bf Variables} & {\bf Definition} & {\bf Reference} \\
\hline
$D$ & the DeFi protocol & \cref{def:3.1} \\
$L$ & the domain-specific language (DSL) & \cref{def:3.1} \\
$S_0$ & the initial and public blockchain state & \cref{def:3.1} \\
$\psi$ & the attack goal & \cref{def:3.1} \\
$P(S_0)$ & the resulting state after executing $P$ on $S_0$ & \cref{def:3.1} \\
$\tilde{P}$ & the attack sketch & \cref{subsec:3.2} \\
$\tilde{P}(S_0)$ & the program state by evaluating $\tilde{P}$ on $S_0$ & \cref{subsec:3.2} \\
$\hole$ & a hole in the attack sketch & \cref{subsec:3.2} \\
$\mu$ & a symbolic representation in $L$ & \cref{subsec:3.2} \\
$\tilde{P}[\mu / \hole]$ & the sketch after filling the hole $\hole$ with $\mu$ & \cref{subsec:3.2} \\
\hline

$\mathbb{F}$ & the set of public DeFi functions accessible to the attacker & \cref{subsec:5.2} \\
$\mathbb{P}$ & the set of AFL operators & \cref{subsec:5.2} \\ 
$\mathbb{T}$ & the set of different tokens appeared in the given DeFi protocol & \cref{subsec:5.2} \\ 
$\mathbb{E}$ & the set of edges in TFG & \cref{subsec:5.2} \\
$\Phi$ & the set of behavioral constraints & \cref{subsec:5.2} \\
$\epsilon$ & the special node in TFG & \cref{subsec:5.2} \\
$\mathbb{G}$ & the token flow graph (TFG) & \cref{subsec:5.2} \\
\hline
$u$ & indicates a token contract & \cref{subsec:5.3} \\
${\sf flow}(u,x,a,b)$ & indicates $x$ amount of token $u$ flows from address $a$ to address $b$ & \cref{subsec:5.3} \\
$\mathbb{W}$ & the flow state & \cref{fig:inf-flow} \\ 
$\mathbb{W} \overset{s}{\rightsquigarrow} \mathbb{W}'$ & the state transition from $\mathbb{W}$ to $\mathbb{W}'$ after executing the statement $s$ & \cref{fig:inf-flow} \\
$u[a]$ & the balance of token $u$ of address $a$ & \cref{fig:inf-flow} \\
$\bullet$ & the dead address & \cref{fig:inf-flow} \\

\hline

$\kappa$ & the knowledge base of constraint clauses & \cref{alg:attack-synthesis} \\
$\delta$ & the constraints of a given sketch & \cref{alg:attack-synthesis} \\
$R$ & a path consisting of a set of edges & \cref{alg:sketch-generation} \\ 
$T$ & the token worklist & \cref{alg:sketch-generation} \\
$\Omega$ & the constraint store as the feedback & \cref{alg:sketch-generation} \\
$\alpha(\psi)$ & the coverage condition & \cref{alg:sketch-generation} \\
${\uparrow}(u, a, x)$ & the constraint of describing a balance increase of $x$ of the token $u$ at address $a$ & \cref{fig:inf-constraints} \\
${\downarrow}(u, a, x)$ & the constraint of describing a balance decrease of $x$ of the token $u$ at address $a$ & \cref{fig:inf-constraints} \\
$p \Downarrow C$ & program $p$ derives the constraint set $C$ by transition rules & \cref{fig:inf-constraints} \\ 
$\varsigma(a, u, v, x, y, b)$ & the constraint describing balance changes between address $a$ and $b$ during swap operation & \cref{fig:inf-constraints} \\
$\rho(x, y)$ & the invariant between the balance $x$ and $y$ of two tokens & \cref{fig:inf-constraints} \\
$\vartheta(x,y)$ & the constraint of additional fee & \cref{fig:inf-constraints} \\

\noalign{\hrule height 1pt}
\end{tabular}
}
\label{tab:notation}
\end{table}

Table~\ref{tab:notation} summarizes the notations used in our paper and their initial appearances. These notations can be categorized into four types:
\begin{enumerate}
\item Notations for defining the attack synthesis process.
\item Notations for describing the token flow graph.
\item Notations for indicating the inference rules of flow predicates.
\item Notations for outlining the domain-specific compilation process.
\end{enumerate}

\end{document}